\documentclass[aps,pre,final]{revtex4}
\usepackage{amsmath,amssymb}
\usepackage{graphicx}
\include{epsf}
\usepackage{clrscode}

\def\(({\left(}
\def\)){\right)}
\def\[[{\left[}
\def\]]{\right]}

\newcommand{\bF}{{\textbf {F}}}
\newcommand{\bG}{{\textbf {G}}}
\newcommand{\bW}{{\textbf {W}}}
\newcommand{\bx}{{\textbf {x}}}
\newcommand{\bo}{{\textbf {o}}}
\newcommand{\bs}{{\textbf {s}}}
\newcommand{\by}{{\textbf {y}}}

\newcommand{\be}{\begin{equation}}
\newcommand{\ee}{\end{equation}}
\newcommand{\bea}{\begin{eqnarray}}
\newcommand{\eea}{\end{eqnarray}}

\newcommand{\ox}{\overline{x}}

\newcommand{\eps}{\epsilon}
\newcommand{\cV}{{\mathcal{V}}}
\def\expect{\mathbb E}

\def\ind{{\mathbb I}}

\begin{document}
\title{Statistical-physics-based reconstruction in compressed sensing}

\author{F. Krzakala $^{1}$, M. M\'ezard $^2$, F. Sausset $^2$, Y. F. Sun$^{1,3}$ and L. Zdeborov\'a $^4$}

\affiliation{ $^1$ CNRS and ESPCI ParisTech, 10 rue Vauquelin, UMR
  7083 Gulliver, Paris 75005, France. \\
$^2$ Univ. Paris-Sud \& CNRS, LPTMS, UMR8626,  B\^{a}t.~100, 91405
Orsay, France. \\
$^3$ LMIB and School of Mathematics and Systems Science, Beihang
University, 100191 Beijing, China. \\
$^4$ Institut de Physique Th\'eorique, IPhT, CEA Saclay, and URA 2306, CNRS, 91191 Gif-sur-Yvette, France.}

\email[Corresponding author; ]{fk@espci.fr}

\begin{abstract}
  Compressed sensing is triggering a major evolution in signal
  acquisition. It consists in sampling a sparse signal at low rate and
  later using computational power for its exact reconstruction, so
  that only the necessary information is measured. Currently used
  reconstruction techniques are, however, limited to acquisition rates
  larger than the true density of the signal.  We design
  a new procedure which is able to reconstruct exactly the signal with
  a number of measurements that approaches the theoretical limit in
  the limit of large systems. It is based on the joint use of three
  essential ingredients: a probabilistic approach to signal
  reconstruction, a message-passing algorithm adapted from belief
  propagation, and a careful design of the measurement matrix inspired
  from the theory of crystal nucleation. The performance of this new
  algorithm is analyzed by statistical physics methods. The obtained 
  improvement is confirmed by numerical studies of several cases.
\end{abstract}

%
\date{\today}
\maketitle

\begin{figure}[ht]
  \begin{center}
    \includegraphics[scale=0.5]{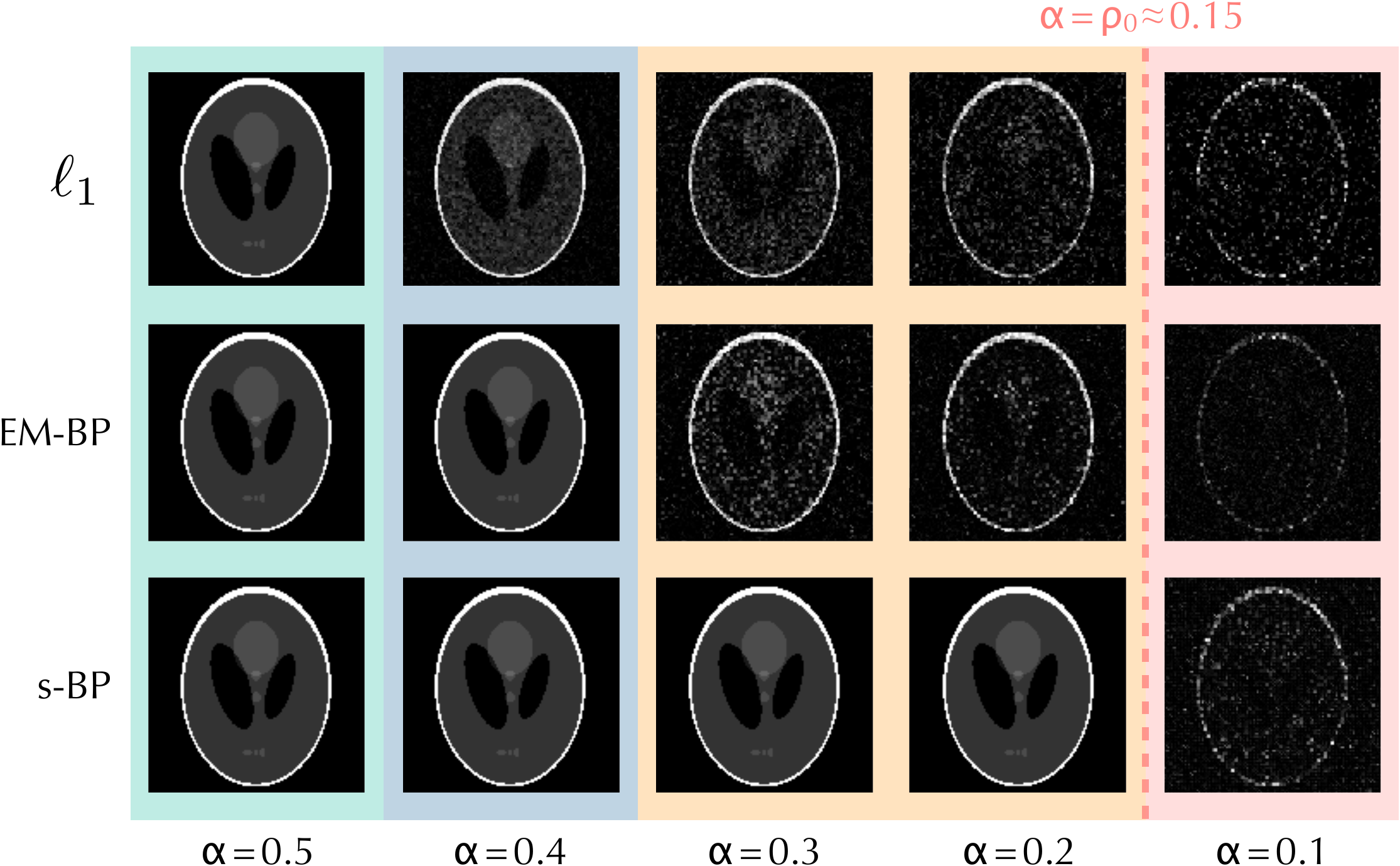}\\[0.2cm]
    \includegraphics[scale=0.5]{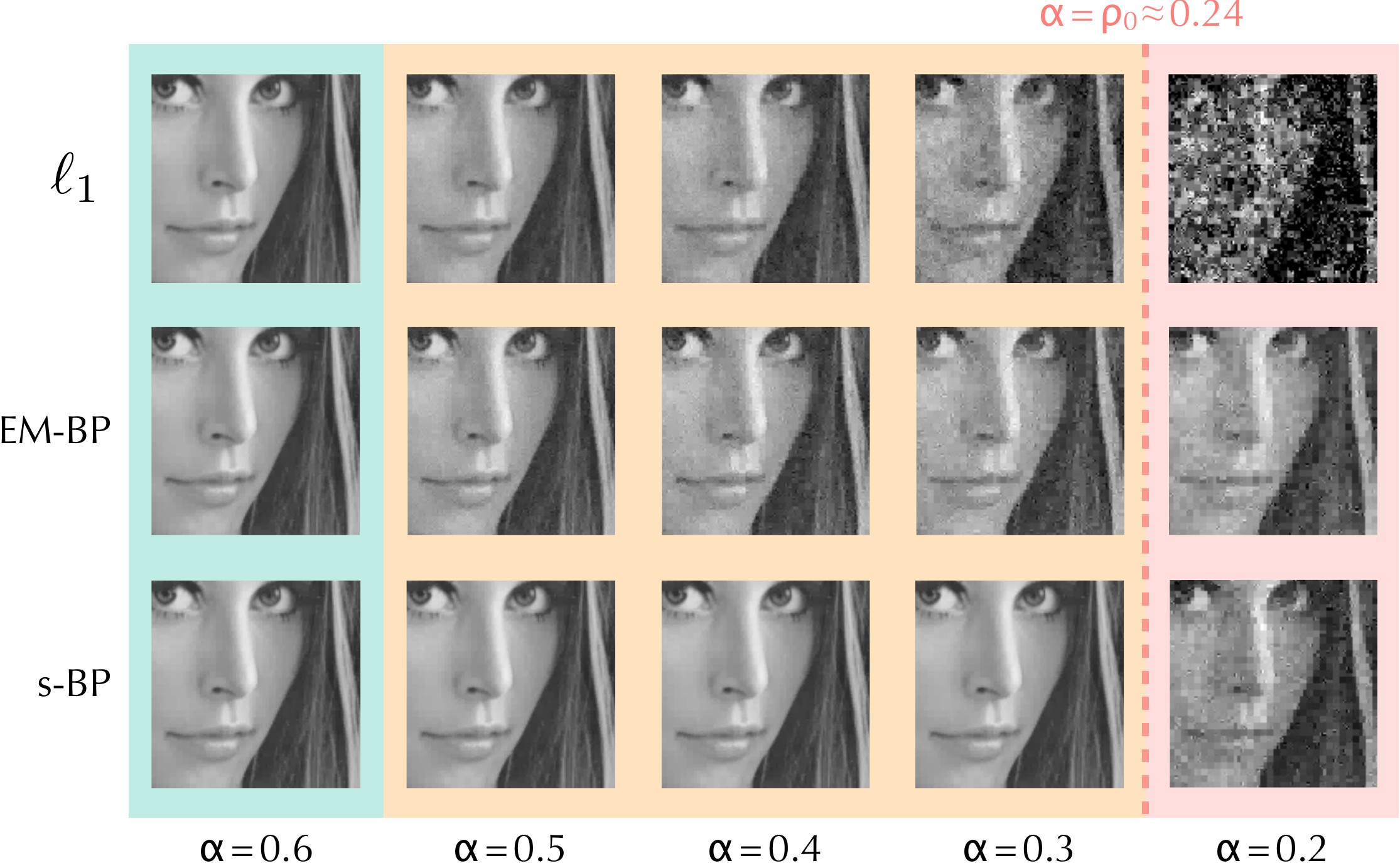}
    \caption{Two illustrative examples of compressed sensing in image
      processing. Top: the original image, the Shepp-Logan phantom, of
      size $N=128^2$, is transformed via one step of Haar wavelets
      into a signal of density $\rho_0\approx 0.15$. With compressed
      sensing one is thus in principle able to reconstruct exactly the
      image with $M \ge\rho_0 N$ measurements, but practical
      reconstruction algorithms generally need $M$ larger than $\rho_0
      N$. The five columns show the reconstructed figure obtained from
      $M=\alpha N$ measurements, with decreasing acquisition rate
      $\alpha$. The first row is obtained with the
      $\ell_1$-reconstruction algorithm~\cite{CandesTao:05,Donoho:06}
      for a measurement matrix with iid elements of zero mean and
      variance $1/N$. The second row is obtained with
      belief propagation, using exactly the same measurements as used
      in the $\ell_1$ reconstruction. The third row is the result of
      the seeded belief propagation, introduced in this work, which
      uses a measurement matrix based on a chain of coupled blocks,
      see Fig.~\ref{fig:1d_matrix}. The running time of all the three
      algorithms is comparable (asymptotically they are all quadratic
      in the size of the signal). In the second part of the figure we
      took as the sparse signal the relevant coefficients after
      two-step Haar transform of the picture of Lena. Again for this
      signal, with density $\rho_0=0.24$, the s-BP procedure
      reconstructs exactly down to very low measurement rates (details
      are in Appendix G, the data is available online
      \cite{noteASPICS}).  }
  \label{fig:phantom}
\end{center}
\end{figure}

The ability to recover high dimensional signals using only a limited
number of measurements is crucial in many fields, ranging from image
processing to astronomy or systems biology. Example of direct
applications include speeding up magnetic resonance imaging without
the loss of resolution, sensing and compressing data simultaneously
\cite{Candes:2008} and the single-pixel camera
\cite{SinglePixel:08}. Compressed sensing  is designed to directly
acquire {\it only} the necessary information about the signal.  This
is possible when the signal is sparse in some known basis. In a {\it second
  step} one uses computational power to reconstruct the signal exactly
\cite{CandesTao:05,Donoho:06,Candes:2008}. Currently, the best known
generic method for exact reconstruction is based on converting the
reconstruction problem into a convex optimization one, which can be
solved efficiently using linear programming
techniques~\cite{CandesTao:05,Donoho:06}. 
The $\ell_1$ reconstruction is able to reconstruct accurately, provided
the system size is large and the ratio of the
number of measurements $M$ to the number of non-zeros $K$ exceeds a speciﬁc limit which
can be proven by careful analysis \cite{Donoho:2005wq,Donoho:06}. However, the limiting
ratio is signiﬁcantly larger than $1$. 
In this paper we improve on the performance of $\ell_1$ minimization, and in the best possible way: we introduce a new procedure that
is able to reach the optimal limit $M/K \to 1$. 
This procedure, which we call {\it seeded Belief
  Propagation} (s-BP) is based on a new, carefully designed, measurement
matrix. It is very powerful, as illustrated in
Fig.~\ref{fig:phantom}. Its performance will be studied here with a
joint use of numerical 
and 
 analytic studies using methods from statistical physics \cite{MezardParisi87b}.

\subsection*{Reconstruction in Compressed Sensing} The mathematical
problem posed in compressed-sensing reconstruction is easily stated. Given an unknown
signal which is a $N$-dimensional vector $\bs$, we make $M$
measurements, where each measurement amounts to a projection of $\bs$
on some known vector. The measurements are grouped into a
$M$-component vector $\by$, which is obtained from $\bs$ by a linear
transformation $\by=\bF \bs$. Depending on the application, this
linear transformation can be for instance associated with measurements
of Fourier modes or wavelet coefficients. The observer knows the
$M\times N$ matrix $\bF$ and the $M$ measurements $\by$, with $M<N$.
His aim is to reconstruct $\bs$. This is impossible in general, but compressed sensing
deals with the case where the signal $\bs$ is sparse, in the sense
that only $K<N$ of its components are non-zero. 
We shall study the case where the non-zero components are real numbers
and the measurements are linearly independent. In this case, 
exact signal reconstruction is possible in principle whenever $M\ge
K+1$, using an exhaustive enumeration method which tries to solve
$\by=\bF \bx$ for all ${N\choose K}$ possible choices of locations of
non-zero components of $\bx$: only one such choice gives a consistent
linear system, which can then be inverted. However, one is typically
interested in large instances where $N\gg 1$, with $M=\alpha N$ and
$K=\rho_0 N$. The enumeration method solves the compressed sensing problem in the
regime where measurement rates are at least as large as the signal
density, $\alpha \ge \rho_0$, but in a time which grows exponentially
with $N$, making it totally impractical.
Therefore $\alpha=\rho_0$ is the fundamental reconstruction limit for
perfect reconstruction in the noiseless case, when the non-zero
components of the signal are real numbers drawn from a continuous distribution.  A  general and
detailed discussion of information-theoretically optimal
reconstruction has been developed recently in
\cite{WuVerdu11b,WuVerdu11,GuoBaron09}.

\begin{figure}[ht]
  \begin{center}
		\includegraphics[scale=0.5]{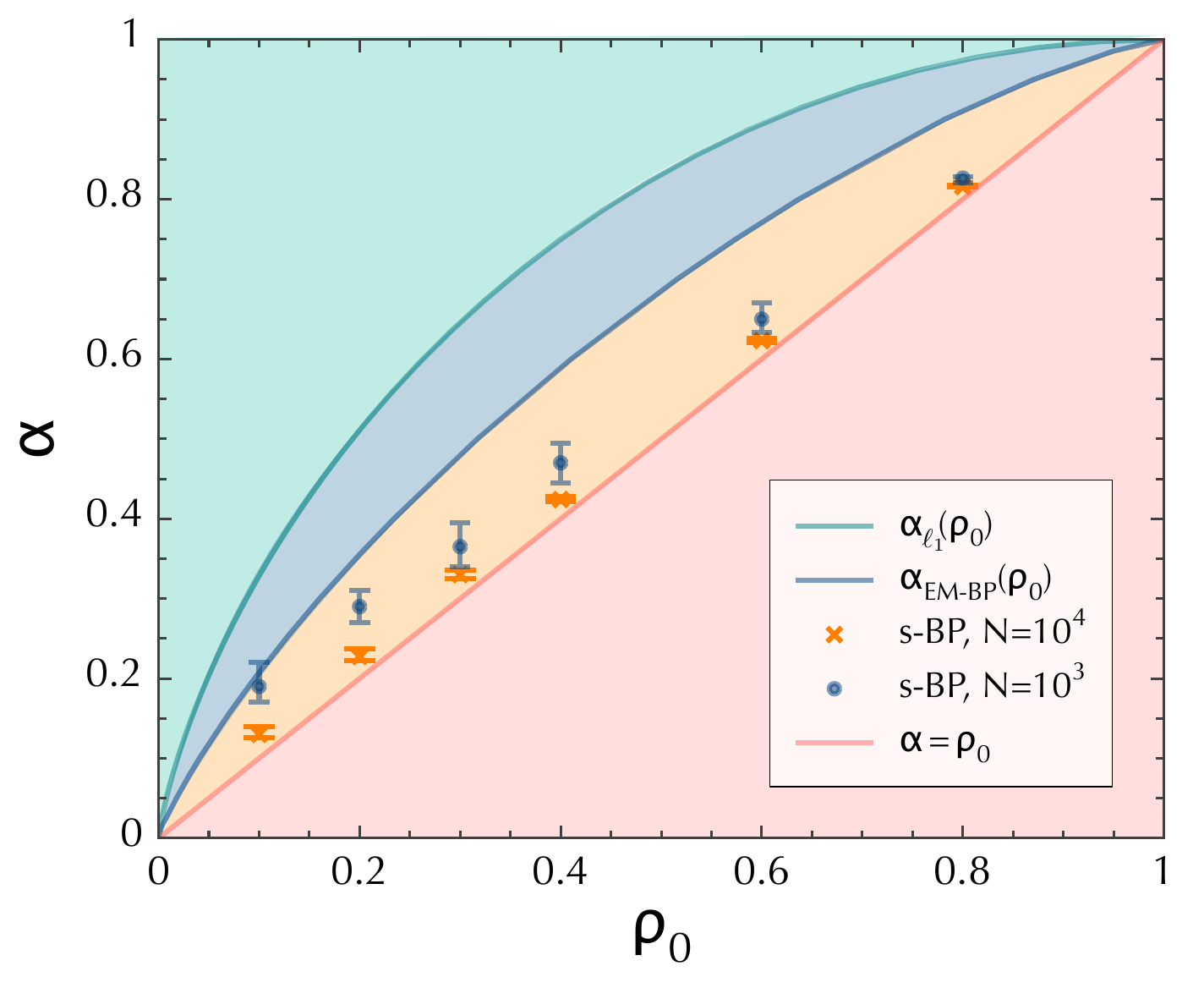}
		\includegraphics[scale=0.5]{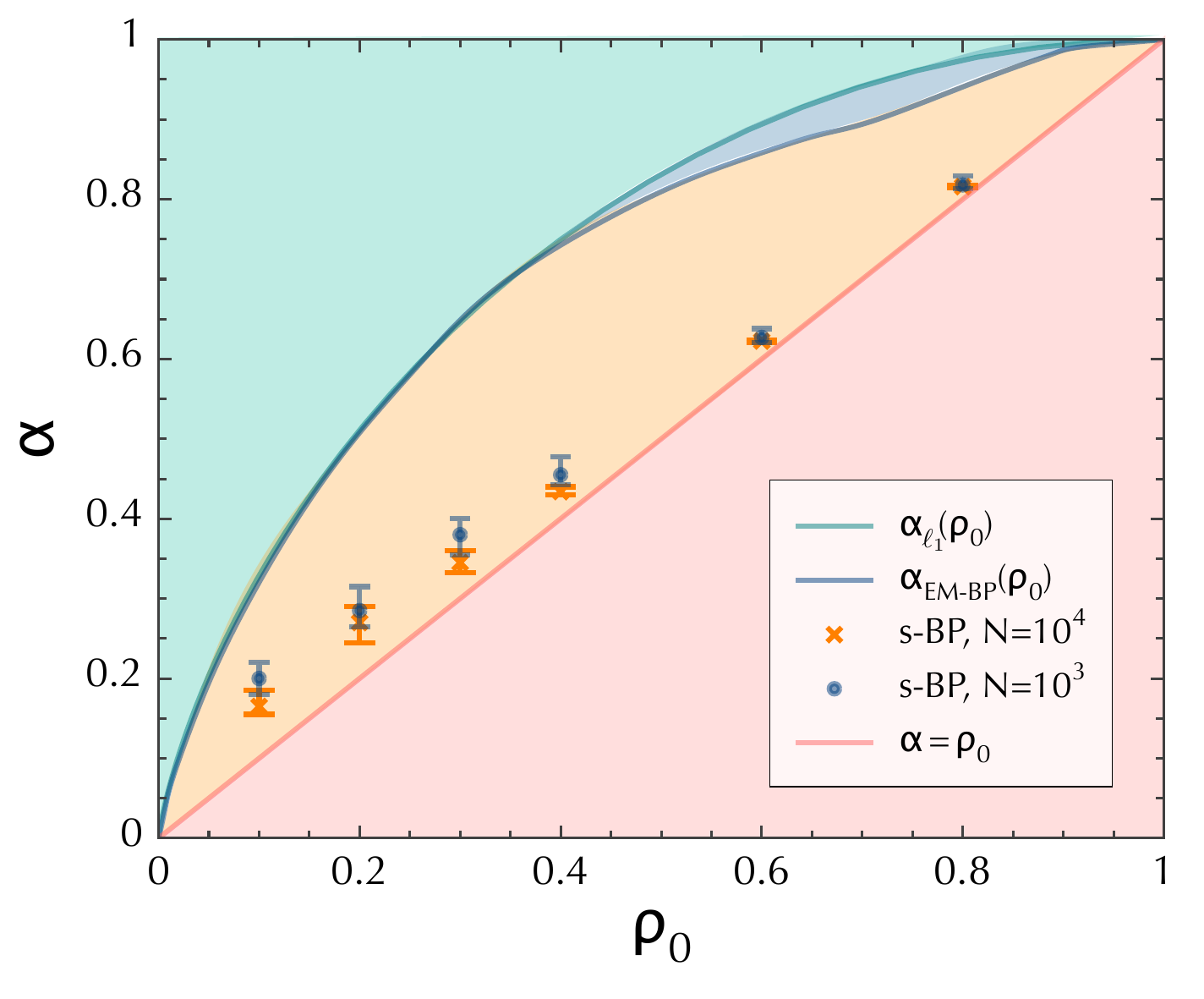}
                \caption{ Phase diagrams for compressed sensing
                  reconstruction for two different signal
                  distributions. On the left-hand side the $\rho_0 N$
                  non-zero components of the signal are independent
                  Gaussian random variables with zero mean and unit
                  variance. On the right-hand side they are
                  independent $\pm 1$ variables.  The measurement rate
                  is $\alpha=M/N$. On both sides we show, from top to
                  bottom: (a) The phase transition $\alpha_{\ell_1}$
                  for $\ell_1$ reconstruction
                  \cite{Donoho:2005wq,DonohoMaleki09,KabashimaWadayama09}
                  (which does not depend on the signal distribution).
                  (b) The phase transition $\alpha_{\rm EM-BP}$ for
                  EM-BP reconstruction based for both sides on the
                  probabilistic model with Gaussian $\phi$. (c) The
                  data points which are numerical reconstruction
                  thresholds obtained with the s-BP procedure with
                  $L=20$. The point gives the value of $\alpha$ where
                  exact reconstruction was obtained in $50$\% of the
                  tested samples, the top of the error bar corresponds
                  to a success rate of $90$\%, the bottom of the bar
                  to a success of $10$\%. The shrinking of the error
                  bar with increasing $N$ gives numerical support to
                  the existence of the phase transition that we have
                  studied analytically. These empirical reconstruction
                  thresholds of s-BP are quite close to the
                  $\alpha=\rho_0$ optimal line, and get closer to it
                  when increasing $N$. The parameters used in these
                  numerical experiments are detailed in Appendix E.
                  (d) The line $\alpha=\rho_0$ that is the theoretical
                  reconstruction limit for signals with continuous
                  $\phi_0$.  An alternative presentation of the same
                  data using the convention of Donoho and Tanner
                  \cite{Donoho:2005wq} is shown in Appendix F.  }
  \label{fig:phase_diag}
\end{center}
\end{figure}

In order to design practical `low-complexity' reconstruction
algorithms, it has been proposed~\cite{CandesTao:05,Donoho:06} to find
a vector $\bx$ which has the smallest $\ell_1$ norm, $\sum_{i=1}^N
|x_i|$, within the subspace of vectors which satisfy the constraints
$\by=\bF \bx$, using efficient linear programming techniques. In order
to measure the performance of this strategy one can focus on the
measurement matrix $\bF$ generated randomly, with independent
Gaussian-distributed matrix elements of mean zero and variance $1/N$,
and the signal vector $\bs$ having density $0<\rho_0<1$. The analytic
study of the $\ell_1$ reconstruction in the thermodynamic limit $N\to
\infty$ can be done using either geometric or probabilistic methods
\cite{Donoho:2005wq,DonohoMaleki09}, or with the replica method
\cite{KabashimaWadayama09,RanganFletcherGoyal09,GanguliSompolinsky10}.
It shows the existence of a sharp phase transition at a value
$\alpha_{\ell_1}(\rho_0)$. When $\alpha$ is larger than this value,
the $\ell_1$ reconstruction gives the exact result $\bx=\bs$ with
probability going to one in the large $N$ limit, when
$\alpha<\alpha_{\ell_1}(\rho_0)$ the probability that it gives the
exact result goes to zero.  As shown in Fig.~\ref{fig:phase_diag},
$\alpha_{\ell_1}(\rho_0)>\rho_0$ and therefore the $\ell_1$
reconstruction is suboptimal: it requires more measurements than would
be absolutely necessary,
in the sense that, if one were willing
to do brute-force combinatorial optimization, no more than $\rho_0 N$ measurements are necessary.

We introduce a new measurement and reconstruction approach, s-BP, that
allows to reconstruct the signal by a practical method, which needs
only $\approx \rho_0 N$ measurements.  We shall now discuss its three
ingredients: 1) a probabilistic approach to signal reconstruction, 2)
a message-passing algorithm adapted from belief
propagation~\cite{Pearl82}, which is a procedure known to be efficient
in various hard computational problems
\cite{RichardsonUrbanke08,MezardMontanari09}, and 3) an innovative
design of the measurement matrix inspired from the theory of crystal
nucleation in statistical physics and from recent developments in
coding theory
\cite{FelstromZigangirov99,KudekarRichardson10,LentmaierFettweis10,HassaniMacris10}.
Some previous works on compressed sensing have used these ingredients
separately.  In particular, adaptations of belief propagation have
been developed for the compressed sensing reconstruction, both in the
context of $\ell_1$ reconstruction
\cite{DonohoMaleki09,DonohoMaleki10,Rangan10b}, and in a probabilistic
approach \cite{Rangan10}. The idea of seeding matrices in compressed
sensing was introduced in \cite{KudekarPfister10}.  It is however only
the combined use of these three ingredients that allows us to reach
the $\alpha = \rho_0$ limit.

\subsection*{A probabilistic approach}
For the purpose of our analysis, we consider the case
where the signal $\bs$ has independent identically distributed (iid)
components: $P_0(\bs) = \prod_{i=1}^N
[(1-\rho_0)\delta(s_i)+\rho_0 \phi_0(s_i)]$, with $0<\rho_0<1$. In the large-$N$ limit
the number of non-zero components  is $\rho_0 N$. Our approach handles
general distributions $\phi_0(s_i)$.

Instead of using a minimization procedure, we shall adopt a
probabilistic approach. We introduce a probability measure $\hat
P(\bx)$ over vectors $\bx \in \mathbb{R}^N$ which is the restriction
of the Gauss-Bernoulli measure $P(\bx)=\prod_{i=1}^N
[(1-\rho)\delta(x_i)+\rho \phi(x_i)]$ to the subspace
$|\by-\bF\bx|=0$~\cite{Note1}. In this paper, we use a distribution
$\phi(x)$ which is a Gaussian with mean $\overline x$ and variance
$\sigma^2$, but other choices for $\phi(x)$ are possible.  It is
crucial to note that we do not require a priori knowledge of the
statistical properties of the signal: we use a value of $\rho$ not
necessarily equal to $\rho_0$, and the $\phi$ that we use is not
necessarily equal to $\phi_0$.  The important point is to use $\rho<1$
(which reflects the fact that one searches a sparse signal).

Assuming that $\bF$ is a random matrix,
 either where all the elements are drawn as independent Gaussian random
 variables with zero mean and the same variance, or of the carefully-designed  type
 of `seeding matrices' described below, we
demonstrate in Appendix A that, for any $\rho_0$-dense original signal
$\bs$, and any $\alpha>\rho_0$ the probability $\hat P(\bs)$ of the
original signal goes to one when $N\to\infty$. This result holds
independently of the distribution $\phi_0$ of the original signal,
which does not need to be known. In practice, we see that $\bs$ also
dominates the measure when $N$ is not very large. In principle,
sampling configurations $\bx$ proportionally to the restricted
Gauss-Bernoulli measure $\hat P(\bx)$ thus gives asymptotically the
exact reconstruction in the whole region $\alpha > \rho_0$. This idea
stands at the roots of our approach, and is at the origin of the
connection with statistical physics (where one samples with the
Boltzmann measure).

\begin{figure}[h]
  \begin{center}
    \includegraphics[scale=0.49]{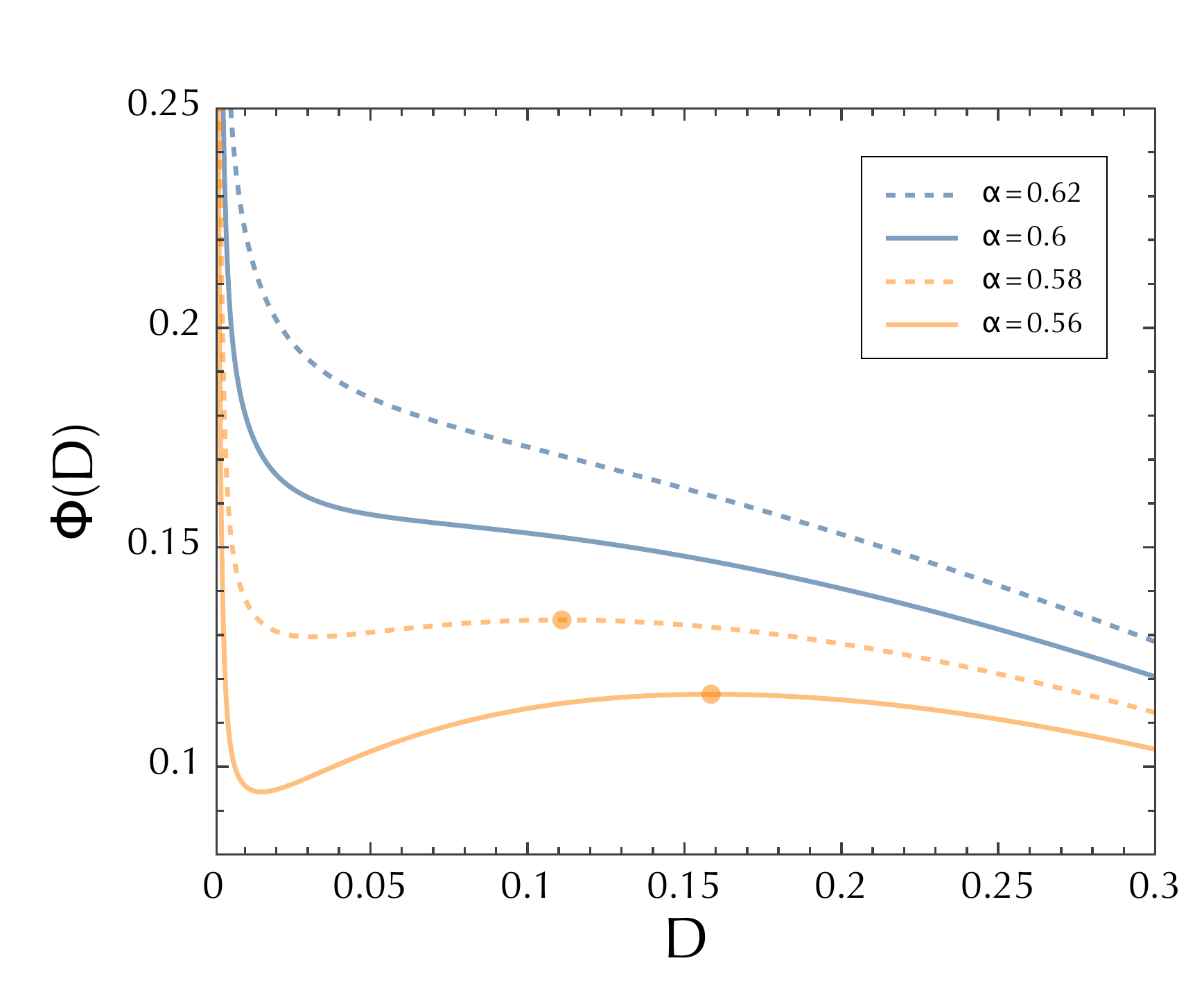}
    \raisebox{0.2cm}{\includegraphics[scale=0.7]{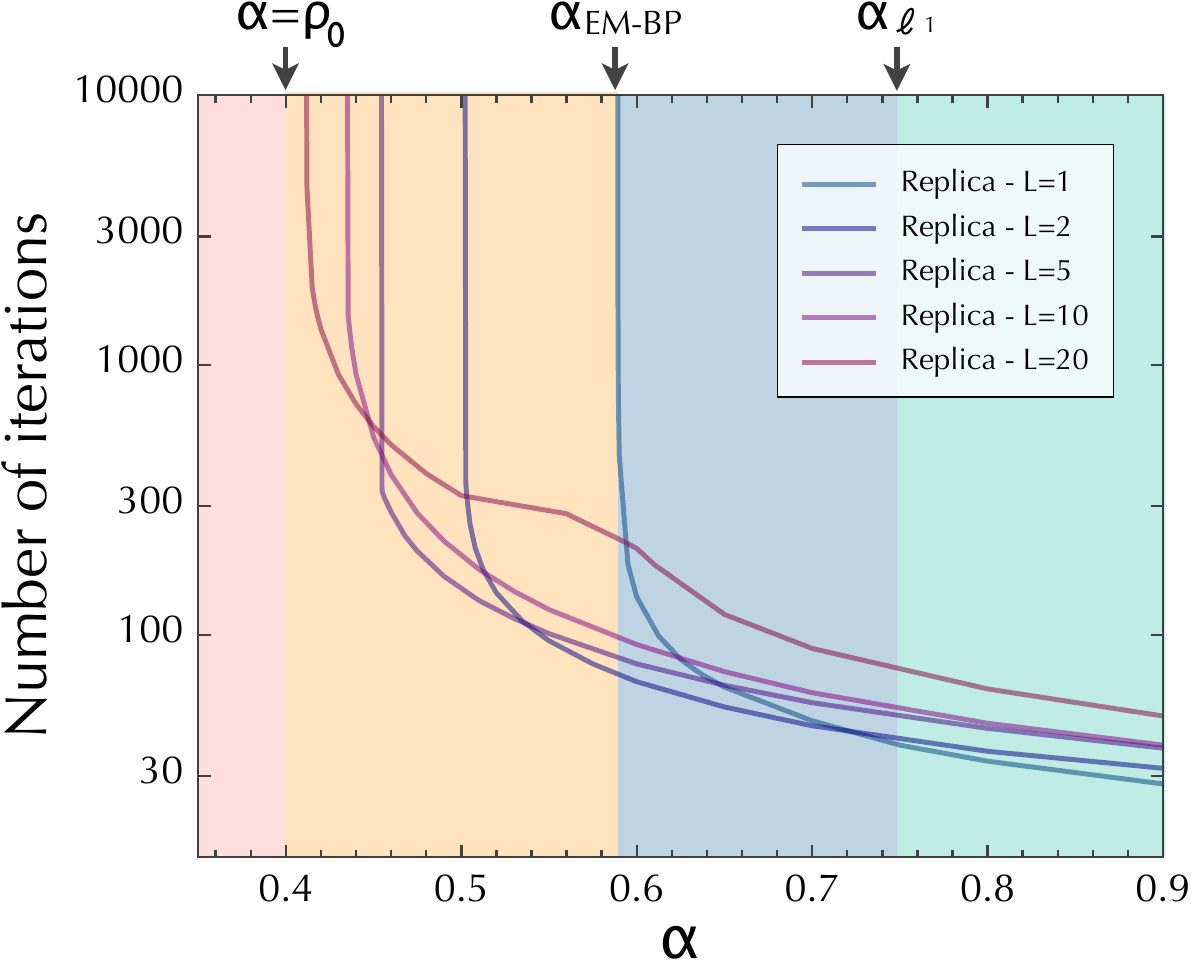}}
    \caption{ When sampling from the probability $\hat P(\bx)$, in the
      limit of large $N$, the probability that the reconstructed
      signal $x$ is at a squared distance $D=\sum_i(x_i-s_i)^2/N$ from
      the original signal $s$ is written as $c e^{N\Phi (D)}$ where
      $c$ is a constant and $\Phi(D)$ is the free entropy. {\bf Left:}
      $\Phi(D)$ for a Gauss-Bernoulli signal with $\rho_0=\rho=0.4$ in
      the case of an unstructured measurement matrix $\bF$ with
      independent random Gaussian-distributed elements. The evolution
      of the EM-BP algorithm is basically a steepest ascent in
      $\Phi(D)$ starting from large values of $D$. It goes to the
      correct maximum at $D=0$ for large value of $\alpha$ but is
      blocked in the local maximum that appears for
      $\alpha<\alpha_{\rm EM-BP}(\rho_0=0.4)\approx0.59$. For
      $\alpha<\rho_0$, the maximum is not at $D=0$ and exact inference
      is impossible. The seeding matrix $\bF$, leading to the s-BP
      algorithm, succeeds in eliminating this local maximum. {\bf
        Right}: Convergence time of the EM-BP and s-BP algorithms
      obtained through the replica analysis for $\rho_0 = 0.4$.  The
      EM-BP convergence time diverges as $\alpha \to \alpha_{\rm
        EM-BP}$ with the standard $L=1$ matrices. The s-BP strategy
      allows to go beyond the threshold: using $\alpha_1=0.7$ and
      increasing the structure of the seeding matrix (here
      $L=2,5,10,20$), we approach the limit $\alpha=\rho_0$ (details
      of the parameters are given in Appendix E).}
   \label{fig:MSQ_BEPL}
  \end{center}
\end{figure}

\subsection*{Sampling with Expectation Maximization Belief Propagation}
The exact sampling from a distribution such as $\hat P(\bx)$, eq.~(\ref{hat_P}), is known to be
computationally intractable~\cite{Natarajan95}.  However, an efficient
approximate sampling can be performed using a message-passing
procedure that we now describe
\cite{ThoulessAnderson77,Tanaka02,GuoWang06,Rangan10}. We start from
the general belief-propagation formalism
\cite{KschischangFrey01,YedidiaFreeman03,MezardMontanari09}: for each
measurement $\mu=1,\dots,M$ and each signal component $i=1,\dots,N$,
one introduces a `message' $m_{i\to \mu}(x_i)$ which is the
probability of $x_i$ in a modified measure where measurement $\mu$ has
been erased.  In the present case,
the canonical belief propagation equations relating these messages can
be simplified
\cite{DonohoMaleki09,DonohoMaleki10,Rangan10,Rangan10b,BayatiMontanari10}
into a closed form that uses only the expectation $a_{i\to \mu}^{(t)}$
and the variance $v_{i\to \mu}^{(t)}$ of the distribution $m_{i\to
  \mu}^{(t)}(x_i)$ (see Appendix B). An
important ingredient that we add to this approach is the learning of
the
parameters in $P(\bx)$: the density $\rho$, and the mean $\overline x$
and variance $\sigma^2$ of the Gaussian distribution $\phi(x)$. These
are three parameters to be learned using update equations based on the
gradient of the so-called Bethe free entropy, in a way analogous to
the expectation
maximization~\cite{Dempster,Iba99,DecelleKrzakala11}. This leads to
the {\it Expectation Maximization Belief Propagation} (EM-BP)
algorithm that we will use in the following for reconstruction in
compressed sensing. It consists in iterating the messages and the
three parameters, starting from random messages $a^{(0)}_{i\to \mu}$
and $v^{(0)}_{i\to \mu}$, until a fixed point is obtained. Perfect
reconstruction is found when the messages converge to the fixed point
$a_{i\to \mu}= s_i$ and $v_{i \to \mu}=0$.

Like the $\ell_1$ reconstruction, the EM-BP reconstruction also has a
phase transition. Perfect reconstruction is
achieved with probability one in the large-$N$ limit if and only if $\alpha>\alpha_{\rm EM-BP}$.   
Using the asymptotic replica analysis, as explained below, we have computed the line $\alpha_{\rm EM-BP}(\rho_0)$
when the elements of the $M\times N$ measurement matrix $\bF$ are
independent Gaussian random variables with zero mean and variance
$1/N$ and the signal components are iid. The location of this
transition line does depend on the signal distribution, see
Fig.~\ref{fig:phase_diag}, contrary to the location of the $\ell_1$ phase
transition. 

Notice that our analysis is fully based on the case when the
probabilistic model has a Gaussian $\phi$. Not surprisingly, EM-BP
performs better when $\phi=\phi_0$, see the left-hand side of
Fig.~\ref{fig:phase_diag}, where EM-BP provides a sizable
improvement over $\ell_1$. In contrast, the right-hand side of
Fig.~\ref{fig:phase_diag} shows an adversary case when we use a
Gaussian $\phi$ to reconstruct a binary $\phi_0$, in this case there
is nearly no improvement over $\ell_1$ reconstruction.

\subsection*{Designing seeding matrices}
In order for the EM-BP message-passing algorithm to be able to
reconstruct the signal down to the theoretically optimal number of
measurements $\alpha=\rho_0$, one needs to use a special family of
measurement matrices $\bF$ that we call `seeding matrices'.  If one
uses an unstructured $\bF$, for instance a matrix with independent
Gaussian-distributed random elements,  EM-BP samples correctly at
large $\alpha$, but at small enough $\alpha$ a metastable state
appears in the measure $\hat P(\bx)$, and the EM-BP algorithm is trapped in
this state, and is therefore unable to find the original signal (see
Fig.~\ref{fig:MSQ_BEPL}),
just as a supercooled liquid gets trapped in a glassy
state instead of crystallizing. It is well known in crystallization
theory that the crucial step is to nucleate a
large enough seed of crystal. This is the purpose of the following
design of $\bF$.

\begin{figure}[h]
  \begin{center}
    \includegraphics[scale=0.45]{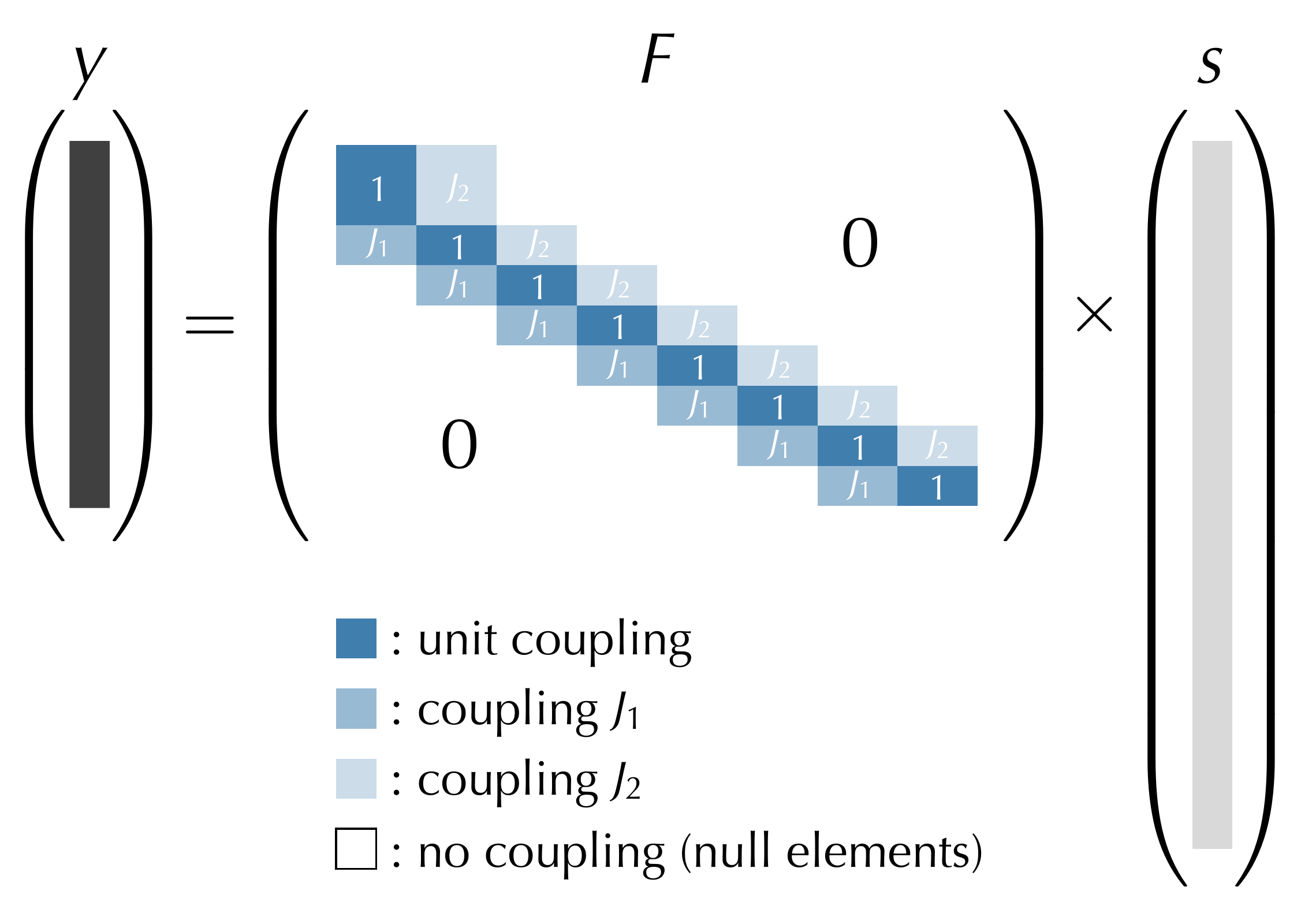}
    \caption{Construction of the measurement matrix $\bF$ for seeded
      compressed sensing. The elements of the signal vector are split
      into $L$ (here $L=8$) equally-sized blocks, the number of
      measurements in each block is $M_p=\alpha_p N/L$
      (here $\alpha_1=1$, $\alpha_p=0.5$ for $p=2,\dots,8$). The
      matrix elements $F_{\mu i}$ are chosen as random Gaussian
      variables with variance $J_{q,p}/N$ if variable $i$ is in the
      block $p$ and measurement $\mu$ in the block $q$.  In the s-BP
      algorithm we use $J_{p,q}=0$, except for $J_{p,p}=1$,
      $J_{p,p-1}=J_1$, and $J_{p-1,p}=J_2$. Good performance is
      typically obtained with relatively large $J_1$ and small $J_2$.}
    \label{fig:1d_matrix}
  \end{center}
\end{figure}

We divide the $N$ variables into $L$ groups of $N/L$ variables, and
the $M$ measurements into $L$ groups. The number of measurements in the $p$-th group is
$M_p=\alpha_p N/L$, so that $M = [(1/L) \sum_{p=1}^L \alpha_p]\; N=
\alpha\; N$. We then choose the matrix elements $F_{\mu i}$
independently, in such a way that, if $i$ belongs to group $p$ and
$\mu$ to group $q$ then $F_{\mu i}$ is a random number chosen from the
normal distribution with mean zero and variance $J_{q,p}/N$ (see
Fig.~\ref{fig:1d_matrix}). The matrix $J_{q,p}$ is a $L\times L$
coupling matrix (and the standard compressed sensing matrices are obtained using $L=1$
and $\alpha_1=\alpha$). Using these new matrices, one can shift the
BP phase transition very close to the
theoretical limit. In order to get an efficient
reconstruction with message passing, one should use a large enough
$\alpha_1$. With a good choice of the coupling matrix $J_{p,q}$, the
reconstruction first takes place in the first block, and propagates as
a wave in the following blocks $p=2,3,\dots$, even if their
measurement rate $\alpha_p$ is small. In practice, we use
$\alpha_2=\dots=\alpha_L=\alpha'$, so that the total measurement rate
is $\alpha=[\alpha_1+(L-1) \alpha']/L$. The whole reconstruction
process is then analogous to crystal nucleation, 
where a crystal is growing from its seed (see Fig.~\ref{fig:surf}). Similar ideas have
been used recently in the design of sparse coding matrices for error-correcting
codes~\cite{FelstromZigangirov99,KudekarRichardson10,LentmaierFettweis10,HassaniMacris10}.

\subsection*{Analysis of the performance of the seeded Belief
  Propagation procedure}
The s-BP procedure is based on the joint use of seeding measurement
matrices and of the EM-BP message-passing reconstruction. We have
studied it with two methods: direct numerical simulations and
analysis of the performance in the large $N$ limit. The
analytical result was obtained by a combination of the replica method
and of the cavity method (also known as `density evolution' or
`state evolution').  The replica method is a standard method in
statistical physics~\cite{MezardParisi87b}, which has been applied
successfully to several problems of information
theory~\cite{NishimoriBook,Tanaka02,GuoVerdu05,MezardMontanari09}
including compressed sensing
\cite{RanganFletcherGoyal09,KabashimaWadayama09,GanguliSompolinsky10}. It
can be used to compute the free entropy function $\Phi$ associated
with the probability $\hat P(\bx)$ (see Appendix D), and
the cavity method shows that the dynamics of the message-passing
algorithm is a gradient dynamics leading to a maximum of this
free-entropy.

When applied to the usual case of the full $\bF$ matrix with
independent Gaussian-distributed elements (case $L=1$), the replica
computation shows that the free-entropy $\Phi(D)$ for configurations
constrained to be at a mean-squared distance $D$  has a global maximum at
$D=0$ when $\alpha > \rho_0$, which confirms that the Gauss-Bernoulli
probabilistic reconstruction is in principle able to reach the optimal
compression limit $\alpha=\rho_0$. However, for $\alpha_{\rm
 EM-BP}>\alpha>\rho_0$, where $\alpha_{\rm EM-BP}$ is a threshold that
depends on the signal and on the distribution $P(\bx)$, a secondary
local maximum of $\Phi(D)$ appears at $D>0$ (see
Fig.~\ref{fig:MSQ_BEPL}). In this case the EM-BP
algorithm converges instead to this secondary maximum and does not
reach exact reconstruction. The threshold $\alpha_{\rm EM-BP}$ is
obtained analytically as the smallest value of $\alpha$ such that
$\Phi(D)$ is 
decreasing  
(Fig.~\ref{fig:phase_diag}).  This theoretical study has been confirmed by
numerical measurements of the number of iterations needed for EM-BP to
reach its fixed point (within a given accuracy).  This convergence
time of BP to the exact reconstruction of the signal diverges when
$\alpha \to \alpha_{\rm EM-BP}$ (see Fig.~\ref{fig:MSQ_BEPL}). For
$\alpha< \alpha_{\rm EM-BP}$ the EM-BP algorithm converges to a fixed point
with strictly positive mean-squared error (MSE). This `dynamical' transition
is similar to the one found in the cooling of liquids which go into a
super-cooled glassy state instead of crystallizing, and appears in the
decoding of error correcting codes
\cite{RichardsonUrbanke08,MezardMontanari09} as well.

We have applied the same technique to the case of seeding-measurement
matrices ($L>1$). The cavity method allows to analytically locate the
dynamical phase transition of s-BP. In the limit of large $N$, the
MSE $E_p$ and the variance messages $V_p$ in each block
$p=1,\dots L$, the density $\rho$, the mean $\overline x$, and the
variance $\sigma^2$ of $P(\bx)$ evolve according to a dynamical system
which can be computed exactly (see Appendix E), and one can
see numerically if this dynamical system converges to the fixed point
corresponding to exact reconstruction ($E_p=0$ for all $p$). This
study can be used to optimize the design of the seeding matrix $\bF$
by choosing $\alpha_1$, $L$ and $J_{p,q}$ in such a way that the
convergence to exact reconstruction is as fast as possible. 
In Fig.~\ref{fig:MSQ_BEPL} we show the convergence time of s-BP
predicted by the replica theory for different sets of parameters. For
optimized values of the parameters, in the limit of a large number of
blocks $L$, and large system sizes $N/L$, s-BP is capable of exact
reconstruction close to the smallest possible number of measurements,
$\alpha\to \rho_0$.
In practice, finite size effects slightly degrade this asymptotic
threshold saturation, but the s-BP algorithm nevertheless reconstructs
signals at rates close to the optimal one regardless of the signal
distribution, as illustrated in Fig.~\ref{fig:phase_diag}.

\begin{figure}[h]
  \begin{center}
    \includegraphics[scale=0.55]{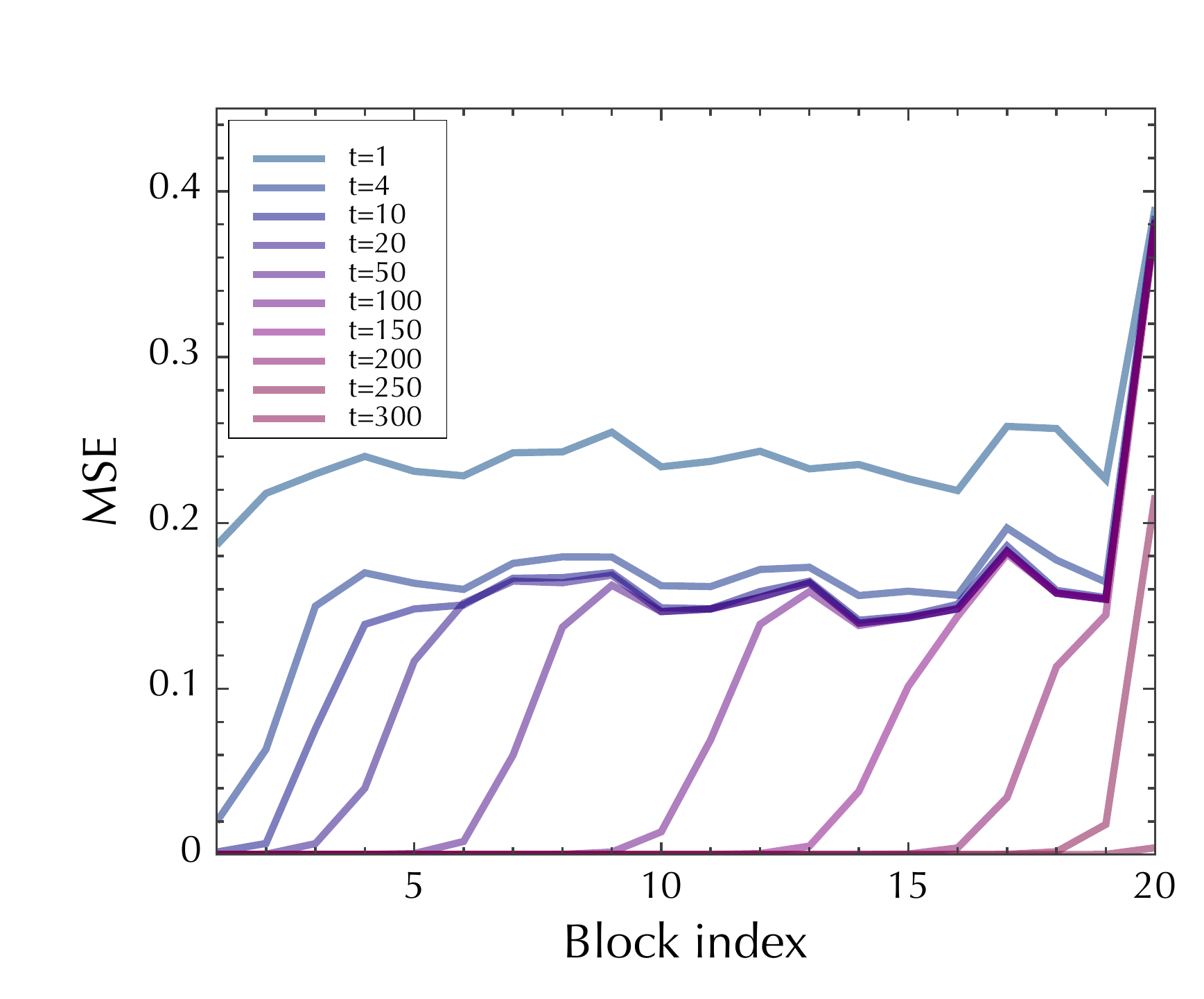}
    \caption{Evolution of the mean-squared error at different times as
      a function of the block index for the s-BP algorithm. Exact
      reconstruction first appears in the left block whose rate
      $\alpha_1>\alpha_{\rm EM-BP}$ allows for seeded nucleation. It
      then propagates gradually block by block driven by the
      free entropy difference between the metastable and equilibrium
      state. After little more than $300$ iterations the whole signal
      of density $\rho_0=0.4$ and size $N=50000$ is exactly
      reconstructed well inside the zone forbidden for BP (see
      Fig.~\ref{fig:phase_diag}). Here we used $L=20$, $J_1 = 20$,
      $J_2 = 0.2$, $\alpha_1=1.0$, $\alpha=0.5$.  }
    \label{fig:surf}
  \end{center}
\end{figure}

We illustrate the evolution of the s-BP algorithm in
Fig.~\ref{fig:surf}. The
non-zero signal elements are Gaussian with zero mean, unit variance,
density $\rho_0=0.4$ and measurement rate $\alpha=0.5$, which is deep in the glassy region where all
other known algorithms fail. The s-BP algorithm
first nucleates the native state in the first block and then
propagates it through the system.  We have also tested the s-BP
algorithm on real images where the non-zero components of the signal
are far from Gaussian, and the results are nevertheless very good, as
shown in Fig.~\ref{fig:phantom}. This shows that the quality of the
result is not due to a good guess of
$P(\bx)$. It is also important to mention that the gain in performance
in using seeding-measurement matrices is really specific to the probabilistic
approach: we have computed the phase diagram of $\ell_1$ minimization
with these matrices and found that, in general, the performance is
slightly degraded with respect to the one of the full measurement
matrices, in the large $N$ limit. This demonstrates that
it is the combination of the probabilistic approach, the
message-passing reconstruction with parameter learning, and the
seeding design of the measurement matrix that is able to reach the best
possible performance.

\subsection*{Perspectives}
The seeded compressed sensing approach introduced here is versatile enough to allow for
various extensions. One aspect worth mentioning is the possibility to
write the EM-BP equations in terms of $N$ messages instead of the
$M\times N$ parameters as described in Appendix C. This is basically
the step that goes from rBP \cite{Rangan10} to AMP
\cite{DonohoMaleki09} algorithm. It could be particularly useful when
the measurement matrix has some special structure, so that the
measurements $\by$ can be obtained in many fewer than $M\times N$
operations (typically in $N\log N$ operations). We have also checked
that the approach is robust to the introduction of a small amount of
noise in the measurements (see Appendix H). Finally, let us mention
that, in the case where a priori information on the signal is
available, it can be incorporated in this approach through a better
choice of $\phi$, and considerably improve the performance of the
algorithm. For signal with density $\rho_0$, the worst case, that we
addressed here, is when the non-zero components of the signal are
drawn from a continuous distribution. Better performance can be
obtained with our method if these non-zero components come from a
discrete distribution and one uses this distribution in the choice of
$\phi$.  Another interesting direction in which our formalism can be
extended naturally is the use of non-linear measurements and different
type of noises.  Altogether, this approach turns out to be very
efficient both for random and structured data, as illustrated in
Fig.~\ref{fig:phantom}, and offers an interesting perspective for
fpractical compressed sensing applications.  Data and code are
available online at \cite{noteASPICS}.

\bf{Acknowledgements}\rm \ We thank Y. Kabashima, R. Urbanke and
specially A. Montanari for useful discussions. This work has been
supported in part by the EC grant `STAMINA', No 265496, and by the
grant DySpaN of `Triangle de la Physique'.

{\bf Note:} During the review process for our paper, we became aware
of the work \cite{DonohoJavanmard11} in which the authors give a
rigorous proof of our result, in the special case when $\rho_0=\rho$
and $\phi_0=\phi$, that the threshold $\alpha=\rho_0$ can be reached
asymptotically by the s-BP procedure.

\newpage
\appendix

\section{Proof of the optimality of the probabilistic approach}
\label{proof}

Here we give the main lines of the proof that our probabilistic
approach is asymptotically optimal. 
We consider the case
where the signal $\bs$ has iid
components
\be 
 P_0(\bs) = \prod_{i=1}^N
[(1-\rho_0)\delta(s_i)+\rho_0 \phi_0(s_i)] \, ,
\ee
with $0<\rho_0<1$. 
And we study the probability distribution 
\be
\hat P(\bx)= 
\frac{1}{Z} \prod_{i=1}^N  \left( dx_i \;
  \left[(1-\rho)\delta(x_i)+\rho \phi(x_i)\right]\right)
\prod_{\mu=1}^M \delta_\epsilon\left(\sum_i F_{\mu
    i}(x_i-s_i)\right) \ , \label{hat_P}
\ee
with a Gaussian $\phi(x)$ of mean zero and unit variance. We stress
here that we consider general $\phi_0$, i.e.  $\phi_0$ is not
necessarily equal to $\phi(x)$ (and $\rho_0$ is not necessarily equal
to $\rho$). The measurement matrix $\bF$ is
composed of iid elements $F_{\mu i}$ such that if $\mu$ belongs to block $q$
and $i$ belongs to block $p$ then $F_{\mu i}$ is  a random number generated from the Gaussian distribution
with zero mean and variance $J_{q,p}$/N. The function $\delta_{\epsilon}(x)$ is a centered Gaussian distribution
with variance $\epsilon^2$.

We show that, with probability going to one in the large $N$ limit (at
fixed $\alpha=M/N$), 
the measure $\hat P$ (obtained with a generic seeding matrix $\bF$ as
described in the main text) is dominated by the signal if
$\alpha>\rho_0$, $\alpha'>\rho_0$ (as long as $\phi_0(0)$
is finite).

We introduce the constrained partition function:
\be
Y(D,\eps)=\int \prod_{i=1}^N  \left( dx_i \;
  \left[(1-\rho)\delta(x_i)+\rho \phi(x_i)\right]\right)
\prod_{\mu=1}^M \delta_\epsilon\left(\sum_i F_{\mu
    i}(x_i-s_i)\right)
\ind \left(\sum_{i=1}^N (x_i-s_i)^2> N D\right) \ ,
\ee
and the corresponding `free entropy density': ${\cal Y}(D,\eps)=\lim_{N\to\infty}\expect_{\bF}\expect_{\bs} \log Y(D,\eps)/N$. The notations $\expect_\bF$ and $\expect_{\bs}$
denote respectively the expectation value with respect to $\bF$ and to
$\bs$. $\ind$ denotes an indicator function, equal to one if its argument is true, and equal to zero otherwise.

The proof of optimality is obtained by showing that, under the
conditions above, $\lim_{\eps \to 0}
{\cal Y}(D,\eps)/[(\alpha-\rho_0)\log(1/\epsilon)]$ is finite if
$D=0$ (statement 1), and it vanishes if $D>0$ (statement 2).
This proves that the measure $\hat P$ is dominated by $D=0$, i.e. by
the neighborhood of the signal $x_i=s_i$. The standard
`self-averageness' property, which states that the distribution (with
respect to the choice of $\bF$ and $\bs$)  of $\log
Y(D,\eps)/N$  concentrates around ${\cal Y}(D,\eps)$ when $N\to\infty$,
completes the proof. We give here the main lines of the first two
steps of the proof.

We first sketch the proof of statement 2. The fact that  $\lim_{\eps \to 0}
{\cal Y}(D,\eps) / [(\alpha-\rho_0)\log(1/\epsilon)] =0$ when $D>0$ can be derived by a first moment bound:
\be
{\cal Y}(D,\eps)\leq \lim_{N\to\infty} \frac{1}{N} \expect_{\bs} \log Y_{\rm ann}(D,\eps)\, ,
\ee
where $Y_{\rm ann}(D,\eps)$ is the `annealed partition function' defined as
\be
Y_{\rm ann}(D,\eps) = \expect_{\bF} Y(D,\eps)\, .
\ee
In order to  evaluate $Y_{\rm ann}(D,\eps)  $ one can
first compute the annealed partition function in which the distances
between $x$ and the signal are fixed in each block. More precisely, we
define
\bea
Z(r_1,\cdots,r_L,\epsilon)= \expect_{\bF} \int \prod_{i=1}^N  \left( dx_i \;
  \left[(1-\rho)\delta(x_i)+\rho \phi(x_i)\right]\right)
\prod_{\mu=1}^M \delta_\epsilon\left(\sum_i F_{\mu
    i}(x_i-s_i)\right)
\prod_{p=1}^L\delta
\left( r_p-\frac{L}{N}\sum_{i\in B_p} (x_i-s_i)^2 \right) \nonumber
\eea
 By noticing
that the $ M$  random variables $a_\mu=\sum_i F_{\mu i}(x_i-s_i)$
are independent Gaussian random variables one obtains:
\be
Z(r_1,\cdots,r_L,\epsilon)= \prod_{p=1}^L
\left(2\pi\left[\epsilon^2+\frac{1}{L}\sum_{q=1}^L J_{pq}r_q\right]\right)^{-N\alpha_p/2}
\prod_{p=1}^L e^{(N/L)\psi(r_p)}\ ,
\label{eq:zzres}
\ee
where
\be
\psi(r)=\lim_{n\to\infty}\frac{1}{n} \log\left[ \int \prod_{i=1}^{n} \left( dx_i \;
  \left[(1-\rho)\delta(x_i)+\rho \phi(x_i)\right]\right)
\delta\left(r-\frac{1}{n}\sum_{i=1}^{n} (x_i-s_i)^2\right)\right]
\ee
The behaviour of $\psi(r)$ is easily obtained by standard saddle point
methods. In particular, when $r\to 0$, one has $\psi(r)\simeq
\frac{1}{2}\rho_0\log r$.

Using (\ref{eq:zzres}), we obtain, in the small $\eps$ limit:
\be
\lim_{N\to\infty} \frac{1}{N} \expect_{\bs} \log Y_{\rm ann}(D,\eps) =
\max_{r_1,\cdots,r_L}
\[[
\frac{\rho_0}{2} \frac{1}{L}\sum_{p=1}^L \log r_p
-\frac{1}{L}\sum_{p=1}^L
\frac{\alpha_p}{2}\log\left(\epsilon^2+\frac{1}{L}\sum_{q=1}^L
  J_{pq}r_q
\right)
\]]\ ,
\ee
where the maximum over $r_1,\dots, r_L$ is to be taken under the
constraint $r_1+\dots+r_L>L D$. Taking the limit of $\epsilon \to 0$
with a finite $D$, at least one of the distance $r_p$ must remain
finite. It is then easy to show that
\be
\lim_{N\to\infty} \frac{1}{N} \expect_{\bs} \log Y_{\rm ann}(D,\eps) =
\log(1/\epsilon)\left[\alpha-\rho_0-\frac{1}{L}(2\alpha'-\rho_0)\right]\ ,
\ee
 where $\alpha'$ is the fraction of measurements in blocks $2$ to
 $L$. As $\alpha'>\rho_0$, this is less singular than
 $\log(1/\epsilon)(\alpha-\rho_0) $, which proves statement 2.

On the contrary, when $D=0$, we obtain from the same analysis
\be
\lim_{N\to\infty} \frac{1}{N} \expect_{\bs} \log Y_{\rm ann}(D,\eps) =
\log(1/\epsilon)(\alpha-\rho_0)
\ee
This annealed estimate actually gives the correct scaling at small
$\epsilon$, as can be shown by the following lower bound.
When $D=0$, we  define $\cV_0$ as the subset of
indices $i$ where $s_i=0$, $|\cV_0|=N(1-\rho_0)$, and $\cV_1$ as the subset of
indices $i$ where $s_i\neq 0$, $|\cV_1|=N\rho_0$. We obtain a lower bound on
$Y(0,\epsilon)$ by
substituting $P(\bx)$ by the factors $(1-\rho)\delta(x_i)$ when
$i\in\cV_0$ and
$\rho \phi(x_i)$ when $i\in\cV_1$.  This gives:
\bea \nonumber
Y(0,\epsilon)&>&\exp\{N [(1-\rho_0)\log(1-\rho)+\rho_0 \log
  (\rho)-(\alpha/2)\log(2 \pi)+( \rho_0-\alpha) \log\epsilon)]\}\\
&&\int \prod_{i\in\cV_1}   du_i \; \phi(s_i + \epsilon u_i)
\;
\exp\left(
-\frac{1}{2} \sum_{i,j\in\cV_1} M_{ij} u_i u_j
\right)\, ,
\label{zzlb}
\eea where $M_{ij}=\sum_{\mu=1}^{\alpha N}F_{\mu i}F_{\mu j}$.  The
matrix $M$, of size $\rho_0N \times \rho_0 N$, is a Wishart-like random
matrix. For $\alpha>\rho_0$, generically, its eigenvalues are strictly
positive, as we show below.
 Using this property, one can show that, if $\prod_{i\in \cV_1}
\phi(s_i)>0$, the integral over the variables $u_i$ in (\ref{zzlb})
is strictly positive in the limit $\epsilon\to 0$. The divergence of $
{\cal Y}(0,\eps)$ in the limit $\eps\to 0$ is due to the explicit term
$\exp[N(\rho_0-\alpha)\log\epsilon]$ in Eq.~(\ref{zzlb}).

The fact that all eigenvalues of $M$ are strictly positive is well
known in the case of $L=1$ where the spectrum has been obtained by
Marcenko and Pastur. 
In general, the fact that all the eigenvalues of $M$ are strictly
positive is equivalent to saying that all the lines of the $\alpha N
\times \rho_0 N$ matrix $F$  (which is the restriction of the
measurement matrix to columns with non-zero signal components) are
linearly independent. In the case of seeding matrices with general
$L$, this statement is basically obvious by construction of the
matrices, in the regime where in each block $q$, $\alpha_q>\rho_0$ and
$J_{qq}>0$. A more formal proof can be obtained as follows.
We consider the Gaussian integral
\be Z(v)=\int \prod_{i=1}^n d \phi_i
\exp\left[-\frac{1}{2}\sum_{i,j,\mu} F_{\mu i} F_{\mu j} \phi_i
  \phi_j+\frac{v}{2}\sum_i\phi_i^2\right] \ee This quantity is finite
if and only if $v$ is smaller than the smallest eigenvalue,
$\lambda_{\rm min}$, of $M$. We now compute the annealed average
$Z_{\rm ann}(v)=\expect_{\bF}Z(v)$. If $Z_{\rm ann}(v) $ is finite,
then
the probability that $\lambda_{\rm min}\le v$ goes to zero in the large N limit.
Using methods similar to the one above, one can show that

\be
\frac{2L}{n}\log\expect_{\bF}Z(v)=\max_{r_1,\dots,r_p}\left[
\sum_{p=1}^L (\log r_p+v r_p)-\sum_{p=1}^L
\frac{\alpha_p}{\rho_0}\log\left(1+\frac{1}{L}\sum_q
  J_{pq}r_q\right)\right]
\ee
The saddle point equations
\be
 \frac{1}{r_p}+v=\frac{1}{L}\sum_{q=1}^L \frac{\alpha_q}{\rho_0} \frac{J_{qp}}{1+\frac{1}{L}\sum_s
  J_{qs}r_s}
\ee
have a solution at $v=0$ (and by continuity also at $v>0$ small
enough), when $\frac{1}{L}\sum_{q=1}^L
\frac{\alpha_q}{\rho_0}=\frac{\alpha}{\rho_0} >1$ (it can be found for
instance by iteration). Therefore, $\frac{2L}{n}\log\expect_{\bF}Z(v)$
is finite for some $v>0$ small enough, and therefore $\lambda_{\rm min}>0$.

\section{Derivation of  Expectation maximization Belief Propagation}
\label{BP_der}
In this and the next sections we present the message-passing algorithm
that we used for reconstruction in compressed sensing. In this section
we derive its message-passing form, where $O(NM)$ messages are being
sent between each signal component $i$ and each measurement
$\mu$.
This algorithm was used in  \cite{Rangan10}, where it was called the
{\it relaxed belief propagation}, as an approximate algorithm for the
case of a  sparse measurement matrix $\bF$. In the case that we use here of
a measurement matrix which is not sparse (a finite fraction of the
elements of $\bF$ is non-zero,
and all the non-zero elements scale as $1/\sqrt{N}$),
 the algorithm is asymptotically exact. We show here for completeness
how to derive it. In the next section we then derive asymptotically
equivalent equations that depend only on $O(N)$ messages. In
statistical physics terms, this corresponds to the TAP equations
\cite{ThoulessAnderson77} with the Onsager reaction term, that are
asymptotically equivalent to the BP on fully connected models. In the
context of compressed sensing this form of equations has been used
previously \cite{DonohoMaleki09} and it is called {\it approximate
  message passing} (AMP). In cases when the matrix $\bF$ can be
computed recursively (e.g. via fast Fourier transform), the running
time of the AMP-type message passing is $O(N\log N)$ (compared to the $O(NM)$ for the non-AMP form). Apart for this speed-up, both classes of message passing give the same performance.

We derive here  the message-passing algorithm in the case where
measurements have additive Gaussian noise, the noiseless case limit is easily
obtained in the end.
The posterior probability of $\bx$ after the measurement of $\by$ is given by
\be
    \hat P(\bx) = \frac{1}{Z}  \prod_{i=1}^N \left[ (1-\rho) \delta(x_i) + \rho \phi(x_i) \right] \prod_{\mu=1}^M \frac{1}{\sqrt{2\pi \Delta_\mu}} e^{-\frac{1}{2\Delta_\mu}(y_\mu - \sum_{i=1}^N F_{\mu i}x_i)^2}\, ,
\ee
where $Z$ is a normalization constant (the partition function) and
$\Delta_\mu$ is the variance of the noise in measurement $\mu$. The
noiseless case is recovered in the limit $\Delta_\mu \to 0$. The optimal
estimate, that minimizes the MSE with respect to the original signal
$\bs$,
is obtained from averages of $x_i$ with respect to the probability measure $\hat P(\bx)$. Exact computation of these averages would require exponential time, belief propagation provides a standard approximation. The canonical BP equations for probability measure  $\hat P(\bx)$ read
\bea
    m_{\mu \to i}(x_i) &=& \frac{1}{Z^{\mu \to i}} \int \left[\prod_{j
        (\neq  i)} m_{j\to \mu}(x_j) {\rm d}x_j\right] e^{-\frac{1}{2\Delta_\mu}(\sum_{j\neq i}F_{\mu j}
      x_j + F_{\mu i}x_i - y_\mu)^2} \, , \label{BP_1}\\
    m_{i\to \mu}(x_i) &=& \frac{1}{Z^{i \to \mu}} \left[ (1-\rho) \delta(x_i) + \rho \phi(x_i) \right] \prod_{\gamma \neq \mu} m_{\gamma \to i}(x_i)  \, .  \label{BP_2}
\eea
where $Z^{\mu \to i}$ and $Z^{i \to \mu}$ are normalization factors ensuring that $\int {\rm d}x_i m_{\mu \to i}(x_i) = \int {\rm d}x_i m_{i \to \mu}(x_i) =1$. These are integral equations for probability distributions that are still practically intractable in this form.
We can, however, take advantage of the fact that after proper rescaling the linear system $\by=\bF \bx$ is such a way that elements of $\by$ and $\bx$ are of $O(1)$, the matrix $F_{\mu i}$ has random elements with variance of $O(1/N)$. Using the Hubbard-Stratonovich transformation
\be
   e^{-\frac{\omega^2}{2\Delta}} =\frac{1}{\sqrt{2\pi \Delta}} \int {\rm d}\lambda e^{-\frac{\lambda^2}{2\Delta}+\frac{i\lambda \omega}{\Delta}}
\ee
for $\omega= (\sum_{j\neq i} F_{\mu j}x_j)$ we can simplify eq.~(\ref{BP_1}) as
\be
 m_{\mu \to i}(x_i) = \frac{1}{Z^{\mu \to i}\sqrt{2\pi \Delta_\mu}} e^{-\frac{1}{2\Delta_\mu}(F_{\mu i}x_i - y_\mu)^2} \int {\rm d}\lambda  e^{-\frac{\lambda^2}{2\Delta_\mu}} \prod_{j\neq i} \left[  \int {\rm d}x_j m_{j\to \mu}(x_j) e^{\frac{F_{\mu j} x_j}{\Delta_\mu}(y_\mu - F_{\mu i}x_i + i\lambda)}     \right] \, .
\ee
Now we expand the last exponential around zero because the term $F_{\mu j}$ is small in $N$, we keep all terms that are of $O(1/N)$. Introducing means and variances as new messages
\bea
      a_{i\to \mu } &\equiv&   \int {\rm d}x_i \, x_i \,  m_{i\to \mu}(x_i) \, ,  \label{a_imu}\\
      v_{i\to \mu } &\equiv&   \int {\rm d}x_i \, x^2_i \,  m_{i\to \mu}(x_i) - a^2_{i\to \mu }   \, . \label{v_imu}
\eea
we obtain
\be
     m_{\mu \to i}(x_i) = \frac{1}{Z^{\mu \to i}\sqrt{2\pi \Delta_\mu}} e^{-\frac{1}{2\Delta_\mu}(F_{\mu i}x_i - y_\mu)^2} \int {\rm d}\lambda e^{-\frac{\lambda^2}{2\Delta_\mu}} \prod_{j\neq i} \left[ e^{\frac{F_{\mu j}a_{j\to \mu}}{\Delta_\mu} (y_\mu - F_{\mu i}x_i + i\lambda) + \frac{F^2_{\mu j}v_{j\to \mu}}{2\Delta^2_\mu}(y_\mu - F_{\mu i}x_i + i\lambda)^2}  \right] \, .
\ee
Performing the Gaussian integral over $\lambda$ we obtain
\be
    m_{\mu \to i}(x_i) = \frac{1}{\tilde Z^{\mu \to i}} e^{-\frac{x^2_i}{2}A_{\mu\to i} + B_{\mu \to i} x_i}\, , \quad \quad    \tilde Z^{\mu \to i} = \sqrt{\frac{2\pi}{A_{\mu \to i}}} e^{\frac{B^2_{\mu \to i}}{2A_{\mu \to i}}}\, ,\label{m_mui}
\ee
where we introduced
\bea
     A_{\mu\to i} &=& \frac{F^2_{\mu i}}{\Delta_\mu + \sum_{j\neq i} F^2_{\mu j} v_{j\to \mu}}  \, , \label{A_mu} \\
     B_{\mu \to i} &=& \frac{F_{\mu i}(y_\mu - \sum_{j\neq i} F_{\mu j}a_{j\to \mu})}{\Delta_\mu + \sum_{j\neq i} F^2_{\mu j} v_{j\to \mu}} \, .  \label{B_mu}
\eea
and the normalization $\tilde Z^{\mu \to i}$ contains all the $x_i$-independent factors. The noiseless case corresponds to $\Delta_\mu=0$.

To close the equations on messages $a_{i\to \mu}$ and $v_{i \to \mu}$
we notice that
\be
     m_{i\to \mu}(x_i) = \frac{1}{\tilde Z^{i\to \mu}} \left[ (1-\rho) \delta(x_i) + \rho \phi(x_i) \right] e^{-\frac{x^2_i}{2}\sum_{\gamma\neq \mu}A_{\gamma \to i} + x_i\sum_{\gamma\neq \mu}B_{\gamma \to i}} \, . \label{m_imu}
\ee
Messages $a_{i\to \mu}$ and $v_{i \to \mu}$ are respectively the mean and variance of the probability distribution $m_{i\to \mu}(x_i)$. For general $\phi(x_i)$ the mean and variance (\ref{a_imu}-\ref{v_imu}) will be computed using numerical integration over $x_i$. Eqs.~(\ref{a_imu}-\ref{v_imu}) together with (\ref{A_mu}-\ref{B_mu}) and (\ref{m_imu}) then lead to closed iterative message-passing equations.

In all the specific examples shown here and in the main part of the
paper we used a Gaussian $\phi(x_i)$  with mean $\overline x$ and variance $\sigma^2$. We define two functions
\begin{align}
&f_a(X,Y)=\left[\frac{\rho (Y+\overline x/\sigma^2)}{\sigma(1/\sigma^2+X)^{3/2}}\right]\left[
(1-\rho)e^{-\frac{(Y+\overline x/\sigma^2)^2}{2(1/\sigma^2+X)}+\frac{\overline x^2}{2\sigma^2}}+\frac{\rho}{\sigma(1/\sigma^2+X)^{1/2}}
\right]^{-1} \, , \label{f_a}\\
&f_c(X,Y)=\left[\frac{\rho}{\sigma(1/\sigma^2+X)^{3/2}}\left(1+\frac{(Y+\overline x/\sigma^2)^2}{1/\sigma^2+X}\right)\right]\left[
(1-\rho)e^{-\frac{(Y+\overline x/\sigma^2)^2}{2(1/\sigma^2+X)}+\frac{\overline x^2}{2\sigma^2}}+\frac{\rho}{\sigma(1/\sigma^2+X)^{1/2}}
\right]^{-1}-f^2_a(X,Y)\, . \label{f_c}
\end{align}
Then the closed form of the BP update is
\begin{eqnarray}
a_{i\to \mu}&=&f_a\left(\sum_{\gamma\neq \mu}A_{\gamma \to
    i},\sum_{\gamma\neq \mu}B_{\gamma \to i}\right)\, , \quad \quad a_{i}=f_a\left(\sum_{\gamma}A_{\gamma \to
    i},\sum_{\gamma}B_{\gamma \to i}\right)\, ,
\label{BP_a_closed}  \\
v_{i\to \mu}&=&f_c\left(\sum_{\gamma\neq \mu}A_{\gamma \to
    i},\sum_{\gamma\neq \mu}B_{\gamma \to i}\right) \, , \quad \quad v_{i}=f_c\left(\sum_{\gamma}A_{\gamma \to
    i},\sum_{\gamma}B_{\gamma \to i}\right)\, , \label{BP_v_closed}
\end{eqnarray}
where the $a_i$ and $v_i$ are the mean and variance of the marginal probabilities of variable $x_i$.

As we discussed in the main text the parameters $\rho$, $\overline x$
and $\sigma$ are usually not known in advance. However, their values
can be learned within the probabilistic approach. A standard way to do so
is called expectation maximization~\cite{Dempster}.
One realizes that the partition function
\be
   Z(\rho,\overline x,\sigma)= \int \prod_{i=1}^N {\rm d}x_i  \prod_{i=1}^N \left[ (1-\rho) \delta(x_i) +  \frac{\rho}{\sqrt{2\pi}\sigma} e^{-\frac{(x_i-\overline x)^2}{2\sigma^2}} \right] \prod_{\mu=1}^M \frac{1}{\sqrt{2\pi \Delta_\mu}} e^{-\frac{1}{2\Delta_\mu}(y_\mu - \sum_{i=1}^N F_{\mu i}x_i)^2}\, ,
\ee
is proportional to the probability of the true parameters $\rho_0,\overline s,\sigma_0,$ given the measurement $\by$. Hence to compute the most probable values of parameters one searches for the maximum of this partition function. Within the BP approach the logarithm of the partition function is the Bethe free entropy expressed as\cite{MezardMontanari09}
\be
   F(\rho,\overline x,\sigma)  = \sum_\mu \log{ Z^\mu} + \sum_i \log{Z^i} - \sum_{(\mu i)} \log{Z^{\mu i}}\label{Bethe}
\ee
where
\bea
    Z^i&=& \int {\rm d} x_i \prod_{\mu} m_{\mu \to i}(x_i) \left[ (1-\rho) \delta(x_i) + \frac{\rho}{\sqrt{2\pi}\sigma} e^{-\frac{(x_i-\overline x)^2}{2\sigma^2}} \right]\, .\\
Z^\mu&=& \int \prod_i {\rm d}x_i \prod_i m_{i\to \mu}(x_i) \frac{1}{\sqrt{2\pi \Delta_\mu}} e^{-\frac{(y_\mu - \sum_i F_{\mu i}x_i)^2}{2\Delta_\mu}}\, , \\
Z^{\mu i}&=& \int {\rm d}x_i m_{\mu \to i}(x_i) m_{i \to \mu}(x_i)\, .
\eea

The stationarity conditions of Bethe free entropy (\ref{Bethe}) with
respect to $\rho$ leads to \be \rho = \frac{\sum_i
  \frac{1/\sigma^2+U_i}{V_i +\overline x/\sigma^2}a_i}{\sum_i \left[ 1
    -\rho + \frac{\rho}{\sigma(1/\sigma^2+ U_{i})^{\frac{1}{2}}}
    e^{\frac{(V_{i}+\overline x/\sigma^2)^2}{2(1/\sigma^2+ U_{i})}
      -\frac{\overline x^2}{2\sigma^2} }
  \right]^{-1}} \label{learn_rho}\, .  \ee where $U_i=\sum_{\gamma}
A_{\gamma \to i}$, and $V_i=\sum_{\gamma} B_{\gamma \to
  i}$. Stationarity with respect to $\overline x$ and $\sigma$ gives
\bea \overline x & = & \frac{\sum_i a_i}{\rho \sum_i \left[\rho + (1 - \rho) \sigma(1/\sigma^2+ U_{i})^{\frac{1}{2}}
    e^{-\frac{(V_{i}+\overline x/\sigma^2)^2}{2(1/\sigma^2+ U_{i})}
      +\frac{\overline x^2}{2\sigma^2} }
  \right]^{-1}}\, , \\
\sigma^2
& = & \frac{\sum_i (v_i + a_i^2)}{\rho \sum_i \left[\rho + (1 - \rho) \sigma(1/\sigma^2+ U_{i})^{\frac{1}{2}}
    e^{-\frac{(V_{i}+\overline x/\sigma^2)^2}{2(1/\sigma^2+ U_{i})}
      +\frac{\overline x^2}{2\sigma^2} }
  \right]^{-1}} - \overline x^2\,
. \label{learn_mv} \eea In statistical physics conditions
(\ref{learn_rho}) are known under the name Nishimori conditions
\cite{Iba99,NishimoriBook}.  In the expectation maximization
eqs.~(\ref{learn_rho}-\ref{learn_mv}) they are used iteratively for
the update of the current guess of parameters. A reasonable initial
guess is $\rho_{\rm init.}=\alpha$.  The value of $\rho_0 \overline s$
can also be obtained with a special line of measurement consisting of
a unit vector, hence we assume that given estimate of $\rho$ the
$\overline x=\rho_0 \overline s/\rho$. In the case where the matrix
$\bF$ is random with Gaussian elements of zero mean and variance
$1/N$, we can also use for learning the variance: $\sum_{\mu=1}^M
y^2_\mu/N = \alpha \rho_0 \langle s^2 \rangle=\alpha \rho (\sigma^2 +
\overline x^2)$.

\section{AMP-form of the message passing}
\label{sec:TAP}

In the large $N$ limit, the messages $a_{i\to \mu}$ and $v_{i\to \mu}$
are nearly independent of  $\mu$, but one must be careful to keep the
correcting Onsager reaction terms.
Let us define
\begin{eqnarray}
\omega_\mu&=& \sum_i F_{\mu i} a_{i \to\mu}\, , \quad \quad \gamma_\mu= \sum_i F_{\mu i}^2  v_{i \to\mu}\, , \\
U_i &= &\sum_{\mu}A_{\mu\to i}\, , \quad \quad V_i = \sum_{\mu}B_{\mu\to i}\, ,
\end{eqnarray}
Then we have
\begin{eqnarray}
U_i&=&\sum_\mu \frac{F^2_{\mu i}}{\Delta_\mu + \gamma_\mu - F^2_{\mu
    i} v_{i\to \mu}} \simeq \sum_\mu \frac{F^2_{\mu i}}{\Delta_\mu +
  \gamma_\mu} \, ,\label{TAP_U} \\
V_i&=& \sum_\mu
\frac{F_{\mu i}(y_\mu - \omega_\mu+ F_{\mu i}a_{i\to \mu})}{\Delta_\mu + \gamma_\mu - F^2_{\mu
    i} v_{i\to \mu}}\simeq
\sum_\mu F_{\mu i}
\frac{(y_\mu - \omega_\mu)}{\Delta_\mu + \gamma_\mu}+
f_a\left(U_i,V_i\right)
\sum_\mu F_{\mu i}^2
\frac{1}{\Delta_\mu + \gamma_\mu}\, . \label{TAP_V}
\end{eqnarray}
We now compute $\omega_\mu$
\begin{eqnarray}
a_{i\to \mu}= f_a\left(U_i-A_{\mu\to i},V_i-B_{\mu\to i}\right)\simeq
a_i-A_{\mu\to i}\frac{\partial f_a}{\partial X}\left(U_i,V_i\right)-
B_{\mu\to i}\frac{\partial f_a}{\partial Y}\left(U_i,V_i\right)\, .
\end{eqnarray}
To express $\omega_\mu=\sum_i F_{\mu i} a_{i\to \mu}$, we see
that the first correction term has a contribution in $F^3_{\mu i}$,
and can be safely neglected. On the contrary, the second term has a
contribution in $F^2_{\mu i}$ which one should
keep. Therefore
\begin{eqnarray}
\omega_\mu= \sum_i F_{\mu i} f_a(U_i,V_i)-\frac{
  (y_\mu-\omega_\mu)}{\Delta_\mu +\gamma_\mu} \sum_i F_{\mu i}^2\frac{\partial
  f_a}{\partial Y}\left(U_i,V_i\right) \, .\label{TAP_al}
\end{eqnarray}
The computation of $\gamma_\mu$ is similar, it gives:
\begin{eqnarray}
\gamma_\mu= \sum_i F_{\mu i}^{2} v_i-\sum_i F_{\mu i}^{3}\frac{ (y_\mu-\omega_\mu)}{\Delta_\mu +\gamma_\mu}\frac{\partial f_c}{\partial Y}\left(U_i,V_i\right) \simeq \sum_i F_{\mu i}^{2} f_c(U_i,V_i)  \, . \label{TAP_ga}
\end{eqnarray}

For a known form of matrix $\bF$ these equations can be slightly
simplified further by using the assumptions of the BP approach about
independence of $F_{\mu i}$ and BP messages. This plus a law of large
number implies that for matrix $\bF$ with Gaussian entries of zero
mean and unit variance one can effectively `replace' every $F^2_{\mu
  i}$ by $1/N$ in eqs.~(\ref{TAP_U},\ref{TAP_V}) and
(\ref{TAP_al},\ref{TAP_ga}). This leads, for homogeneous or bloc
matrices, to even simpler equations and a slightly faster algorithm.

Eqs.~(\ref{TAP_U},\ref{TAP_V}) and (\ref{TAP_al},\ref{TAP_ga}), with
or without the later simplification, give a system of closed
equations. They are a special form ($P(\bx)$ and hence functions
$f_a$, $f_c$ are different in our case) of the approximate message
passing of \cite{DonohoMaleki09}.

The final reconstruction algorithm for general measurement matrix and
with learning of the $P(\bx)$ parameters can hence be summarized in a schematic way:
\begin{codebox}
\Procname{$\proc{EM-BP}(y_\mu,F_{\mu i},{\rm criterium},t_{\rm max})$}
\li Initialize randomly messages $U_i$ from interval $[0,1]$ for every component;
\li Initialize randomly messages $V_i$ from interval $[-1,1]$ for every component;
\li Initialize messages $\omega_\mu \gets y_\mu$;
\li Initialize randomly messages $\gamma_\mu$ from interval $]0,1]$ for every measurement;
\li Initialize the  parameters $\rho \gets \alpha$, $\overline x \gets 0$, $\sigma^2 \gets 1$.
\li ${\rm conv} \gets {\rm criterium}+1$; $t \gets 0$;
\li \While ${\rm conv}>{\rm criterium}$ and $t< t_{\rm max}$:
\li     \Do $t \gets t+1$;
\li		\For each component $i$:
\li     	\Do ${x_i^{\rm old}} \gets f_a(U_i, V_i)$;
\li			Update $U_i$ according to eq.~\eqref{TAP_U}.
\li			Update $V_i$ according to eq.~\eqref{TAP_V}.
		\End
\li		\For each measurement $\mu$:
\li			\Do Update $\omega_\mu$ according to eq.~\eqref{TAP_al}.
\li			Update $\gamma_\mu$ according to eq.~\eqref{TAP_ga}.
			\End
\li 	Update $\rho$ according to eq.~\eqref{learn_rho}.
\li 	Update $\overline x$ and $\sigma^2$ according to eq.~\eqref{learn_mv}.
\li		\For each component $i$:
\li     	\Do ${x_i} \gets f_a(U_i, V_i)$;
		\End
\li     ${\rm conv} \gets {\rm mean}(|x_i - x_i^{\rm old}|)$;
    \End
\li \Return signal components $\bx$
\end{codebox}
Note that in practice we use `damping' (at each update, the new
message is obtained as $u$ times the old value plus $1-u$ times the
newly computed value, with a damping $0<u<1$, typically $u=0.5$) for both the update of messages and learning of parameters, empirically this speeds up the convergence.
Note also that the algorithm is relatively robust with respect to the
initialization of messages. The reported initialization was used to
obtain the  results in Fig.~1 of the main text. However, other initializations are possible. Note also that for specific classes of signals or measurement matrices the initial conditions may be adjusted to take into account the magnitude of the values $F_{\mu i}$ and $y_\mu$.

For a general matrix $\bF$ one iteration takes $O(NM)$ steps, we
observed the number of iterations needed for convergence to be
basically independent of $N$, however, the constant depends on the parameters and the signal, see Fig.~3 in the main paper. For matrices that can be computed recursively (i.e. without storing all their $NM$ elements) a speed-up is possible, as the message-passing loop takes only $O(M+N)$ steps.

\section{Replica analysis and  density evolution: full measurement matrix}
\label{replica}
Averaging over disorder leads to replica equations that are describing the $N\to \infty$ behavior of the partition function as well as the density evolution of the belief propagation algorithm. The replica trick evaluates $\mathbb{E}_{{\bf F}, {\bf s}} (\log{Z})$ via
\be
   \Phi = \frac{1}{N} \mathbb{E}(\log{Z}) =\frac{1}{N} \lim_{n\to 0} \frac{\mathbb{E}(Z^n)-1}{n} \, .
\ee

In the case where the matrix $\bF$ is the full measurement with
all elements independent identically distributed from a normal
distribution with zero mean and variance unity, one finds that $\Phi$
is obtained as the saddle point  value of the function:
\bea
     && \Phi(Q,q,m,\hat Q,\hat q,\hat m) = -\frac{\alpha}{2}
     \frac{q-2m+\rho_0 \langle s^2 \rangle+\Delta}{\Delta +Q-q} - \frac{\alpha}{2}  \log{(\Delta+Q-q)} + \frac{Q\hat Q}{2} - m\hat m + \frac{q \hat q}{2} \nonumber \\ && + \int {\cal D}z \int {\rm d}s  \left[(1-\rho_0)\delta(s) + \rho_0 \phi_0(s) \right] \log{\left\{  \int {\rm d}x \, e^{-\frac{\hat Q+\hat q}{2}x^2 + \hat m x s + z \sqrt{\hat q}x} \left[  (1-\rho)\delta(x) +\rho\phi(x) \right] \right\}}\, .\label{free_rep}
\eea
Here ${\cal D}z$ is a Gaussian integration measure with zero mean and
variance equal to one, $\rho_0$ is the density of the signal,
and $\phi_0(s)$ is the distribution of the signal components and $\langle s^2 \rangle= \int ds
s^2 \phi_0(s)$ is its second moment. $\Delta$ is the variance of the
measurement noise, the noiseless case is recovered by using $\Delta=0$.

The physical meaning of the order parameters is
\be
 Q = \frac{1}{N} \sum_i \langle x_i^2 \rangle \, , \quad  q = \frac{1}{N} \sum_i \langle x_i \rangle^2 \, , \quad
 m = \frac{1}{N} \sum_i s_i \langle x_i \rangle \, .
\ee
Whereas the other three $\hat m$, $\hat q$, $\hat Q$ are auxiliary
parameters.
Performing saddle point derivative with respect to $m,q,Q-q,\hat m,
\hat q, \hat Q + \hat q$ we obtain the following six self-consistent
equations  (using the Gaussian form of $\phi(x)$, with mean
$\overline x$ and variance $\sigma^2$):
\bea
   \hat m &=& \frac{\alpha}{\Delta + Q-q}= \hat Q + \hat q  \, , \quad \quad
      \hat Q =  \frac{\alpha}{\Delta + Q-q} - \alpha \frac{q-2m+\rho_0 \langle s^2\rangle+\Delta}{(\Delta + Q-q)^2} \, , \label{hats_gen} \\
     m &=& \frac{\rho_0 \rho}{\hat Q + \hat q + 1/\sigma^2} \int {\cal D}z \int {\rm d} s \, s \phi_0(s) \frac{ \hat m s + z \sqrt{\hat q} + \overline x /\sigma^2   }{(1-\rho) \sigma \sqrt{ \hat Q + \hat q + 1/\sigma^2  } e^{\frac{\overline x^2}{2\sigma^2}-\frac{( \hat m s + z \sqrt{\hat q} + \overline x /\sigma^2  )^2}{2(\hat Q + \hat q + 1/\sigma^2)}} + \rho } \, , \label{eq_m_gen}\\
     Q-q &=& \frac{(1- \rho_0)\rho}{(\hat Q + \hat q + 1/\sigma^2)\sqrt{\hat q}} \int {\cal D}z  \, z \frac{ z \sqrt{\hat q} + \overline x /\sigma^2   }{(1-\rho) \sigma \sqrt{ \hat Q + \hat q + 1/\sigma^2  } e^{\frac{\overline x^2}{2\sigma^2}-\frac{(  z \sqrt{\hat q} + \overline x /\sigma^2  )^2}{2(\hat Q + \hat q + 1/\sigma^2)}} + \rho } \nonumber \\ &+& \frac{\rho_0 \rho}{(\hat Q + \hat q + 1/\sigma^2)\sqrt{\hat q}} \int {\cal D}z \, z\int {\rm d} s \phi_0(s)
\frac{ \hat m s + z \sqrt{\hat q} + \overline x /\sigma^2   }{(1-\rho) \sigma \sqrt{ \hat Q + \hat q + 1/\sigma^2  } e^{\frac{\overline x^2}{2\sigma^2}-\frac{( \hat m s + z \sqrt{\hat q} + \overline x /\sigma^2  )^2}{2(\hat Q + \hat q + 1/\sigma^2)}} + \rho }
   \, , \label{eq_Q_GB}\\
Q  &=& \frac{(1-\rho_0) \rho}{( \hat Q + \hat q + 1/\sigma^2 )^2} \int {\cal D}z \frac{ ( z \sqrt{\hat q} + \overline x /\sigma^2)^2 +  \hat Q + \hat q + 1/\sigma^2  }{(1-\rho) \sigma \sqrt{ \hat Q + \hat q + 1/\sigma^2  } e^{\frac{\overline x^2}{2\sigma^2}-\frac{(  z \sqrt{\hat q} + \overline x /\sigma^2  )^2}{2(\hat Q + \hat q + 1/\sigma^2)}} + \rho } \nonumber \\ &+& \frac{\rho_0 \rho}{( \hat Q + \hat q + 1/\sigma^2 )^2} \int {\cal D}z \int {\rm d} s  \phi_0(s)
\frac{ (\hat m s + z \sqrt{\hat q} + \overline x /\sigma^2)^2 +  \hat Q + \hat q + 1/\sigma^2  }{(1-\rho) \sigma \sqrt{ \hat Q + \hat q + 1/\sigma^2  } e^{\frac{\overline x^2}{2\sigma^2}-\frac{( \hat m s + z \sqrt{\hat q} + \overline x /\sigma^2  )^2}{2(\hat Q + \hat q + 1/\sigma^2)}} + \rho }
     \ .  \label{eq_q_gen}
\eea

We now show the connection between this replica computation and the
evolution of belief propagation messages, studying first the case
where one does not change the parameters $\rho$, $\overline x$ and $\sigma$. Let us introduce parameters  $m_{\rm BP}$, $q_{\rm BP} $, $Q_{\rm BP}$ defined via the belief propagation messages as:
\be
      m^{(t)}_{\rm BP} = \frac{1}{N} \sum_{i=1}^N a^{(t)}_i s_i   \, ,\quad    q^{(t)}_{\rm BP}= \frac{1}{N} \sum_{i=1}^N (a^{(t)}_i)^2 \, ,\quad   Q^{(t)}_{\rm BP}-q^{(t)}_{\rm BP}= \frac{1}{N} \sum_{i=1}^N  v^{(t)}_i\, .
\ee
The density (state) evolution equations for these parameters can be
derived in the same way as in \cite{DonohoMaleki09,BayatiMontanari10},
and this leads to the result that $m_{\rm BP}$, $q_{\rm BP}$, $Q_{\rm
  BP}$ evolve under the update of BP in exactly the same way as
according to iterations of eqs.~(\ref{eq_m_gen}-\ref{eq_q_gen}). Hence
the analytical eqs.~(\ref{eq_m_gen}-\ref{eq_q_gen}) allow to study
the performance of the BP algorithm. Note also that the density
evolution equations are the same for the message-passing and for the AMP equations.
It turns out that the above equations close in terms of two
parameters, the mean-squared error $E^{(t)}_{\rm BP}=q^{(t)}_{\rm
  BP}-2m^{(t)}_{\rm BP}+\rho_0 \langle s^2\rangle$ and the variance
$V^{(t)}_{\rm BP}=Q^{(t)}_{\rm BP}-q^{(t)}_{\rm BP}$. From
eqs.~(\ref{hats_gen}-\ref{eq_q_gen}) easily gets a closed mapping
$\left(E^{(t+1)}_{\rm BP},V^{(t+1)}_{\rm BP}\right)= f \left(E^{(t)}_{\rm BP},V^{(t)}_{\rm BP}\right) $.

In the main text we defined the function $\Phi(D)$ which is the free
entropy restricted to configurations $\bx$ for which $D = \sum_{i=1}^N
(x_i - s_i)^2 /N$ is fixed. This is evaluated as the saddle point over
$Q,q,\hat Q,\hat q,\hat m $ of the function $
\Phi(Q,q,(Q-D+\rho_0\langle s^2 \rangle)/2,\hat Q,\hat q,\hat
m)$. This  function is plotted in Fig.~3(a) of the main text.

In presence of Expectation Maximization learning of the parameters, the density evolution for the conditions (\ref{learn_rho}) and (\ref{learn_mv}) are
\bea
\rho^{(t+1)} &=&\rho^{(t)}
\left(\int {\cal D}z \int {\rm d}x_0
\left[(1-\rho_0)\delta(x_0) +\rho_0 \phi_0(x_0) \right]
\frac{
g(\hat Q+\hat q,\hat m x_0 + z \sqrt{\hat
      q})
}
{
1-\rho+\rho g(\hat Q+\hat q,\hat m x_0 + z \sqrt{\hat
      q})
}
\right) \nonumber \\
&&
\left(\int {\cal D}z \int {\rm d}x_0
\left[(1-\rho_0)\delta(x_0) +\rho_0 \phi_0(x_0) \right]
\frac{
1
}
{
1-\rho+\rho g(\hat Q+\hat q,\hat m x_0 + z \sqrt{\hat
      q})
}
\right)^{-1} \label{eq:learn_analytic1}
\\
\ox^{(t+1)} &=&
\frac{1}{\rho}
\left(
\int {\cal D}z \int {\rm d}x_0
\left[(1-\rho_0)\delta(x_0) +\rho_0 \phi_0(x_0) \right]
f_a(\hat Q+\hat q,\hat m x_0 + z \sqrt{\hat
      q})
\right) \nonumber \\
&&
\left(
\int {\cal D}z \int {\rm d}x_0
\left[(1-\rho_0)\delta(x_0) +\rho_0 \phi_0(x_0) \right]
\frac{
g(\hat Q+\hat q,\hat m x_0 + z \sqrt{\hat q})
}
{
1-\rho+\rho g(\hat Q+\hat q,\hat m x_0 + z \sqrt{\hat
      q})
}
\right)^{-1} \label{eq:learn_analytic2}
\\
(\sigma^2)^{(t+1)}&=&
\frac{1}{\rho}
\left(
\int {\cal D}z \int {\rm d}x_0
\left[(1-\rho_0)\delta(x_0) +\rho_0 \phi_0(x_0) \right]
[f_a(\hat Q+\hat q,\hat m x_0 + z \sqrt{\hat
      q})^2+f_c(\hat Q+\hat q,\hat m x_0 + z \sqrt{\hat
      q})]\right) \nonumber \\
&&
\left(
\int {\cal D}z \int {\rm d}x_0
\left[(1-\rho_0)\delta(x_0) +\rho_0 \phi_0(x_0) \right]
\frac{
g(\hat Q+\hat q,\hat m x_0 + z \sqrt{\hat q})
}
{
1-\rho+\rho g(\hat Q+\hat q,\hat m x_0 + z \sqrt{\hat
      q})
}
\right)^{-1} 
-[\overline x^{(t+1)}]^2
\label{eq:learn_analytic3}
\eea

The density evolution equations now provide a mapping
\be
\left(E^{(t+1)}_{\rm EM-BP},V^{(t+1)}_{\rm EM-BP},\rho^{(t+1)},\overline x^{(t+1)},\sigma^{(t+1)}\right)= f
\left(E^{(t)}_{\rm EM-BP},V^{(t)}_{\rm EM-BP},\rho^{(t)},\overline
  x^{(t)},\sigma^{(t)}\ \right)
\ee  obtained by complementing the
  previous equations on $E^{(t)}_{\rm EM-BP},V^{(t)}_{\rm EM-BP}$ with
  the update equations
  (\ref{eq:learn_analytic1},\ref{eq:learn_analytic2},\ref{eq:learn_analytic3}). The
  next section gives explicitly the full set of equations in the case of seeding
  matrices, the ones for the full matrices are obtained by taking $L=1$.
These are the equations that we study to describe analytically  the
evolution of EM-BP algorithm and obtain the phase diagram for the reconstruction (see Fig. 2 in the main text).

\section{Replica analysis and density evolution for seeding-measurement matrices}
\label{replica_1D}
Many choices of $J_1$ and $J_2$ actually work very well, and good
performance for seeding-measurement matrices can be easily obtained. In
fact, the form of the matrix that we have used is by no means the only one
that can produce the seeding mechanism, and we expect that better
choices, in terms of convergence time, finite-size effects and
sensibility to  noise, could be unveiled in the near future. 

With the matrix presented in this work, and in order to obtain the
best performance (in terms of phase transition limit and of speed of
convergence) one needs to optimize the value of $J_1$ and $J_2$
depending on the type of signal. Fortunately, this can be analysed
with the replica method. The analytic study in the case of
seeding measurement matrices is in fact done using the same techniques
as for the full matrix.  The order parameters are now the MSE
$E_p=q_p-2m_p+\rho_0\langle s^2 \rangle$ and variance$V_p=Q_p-q_p$ in
each block $p\in\{1,\dots,L\}$. Consequently, we obtain the final
dynamical system of $2L+3$ order parameters describing the density
evolution of the s-BP algorithm.  The order parameters at iteration
$t+1$ of the message-passing algorithm are given by:

\bea
  E_q^{(t+1)}  &=& \int {\rm d}x^0
      \left[(1-\rho_0)\delta(x^0)+\rho_0 \phi_0(x^0) \right] \int
      {\cal D}z  \left(
 f_a\left( \hat Q_q+\hat q_q,   \hat m_q x^0 + z \sqrt{\hat
    q_q}\right)-x^0\right)^2\\
V_q^{(t+1)}  & =&\int {\rm d}x^0 \left[(1-\rho_0)\delta(x^0)+\rho_0
     \phi_0(x^0) \right] \int {\cal D}z   f_c\left( \hat Q_q+\hat q_q,   \hat m_q x^0 + z \sqrt{\hat
    q_q}\right)\\
\rho^{(t+1)} &=&\rho^{(t)}
\left(
\frac{1}{L}\sum_{p=1}^L\int {\cal D}z \int {\rm d}x_0
\left[(1-\rho_0)\delta(x_0) +\rho_0 \phi_0(x_0) \right]
\frac{
g(\hat Q_p+\hat q_p,\hat m_p x_0 + z \sqrt{\hat
      q_p})
}
{
1-\rho+\rho g(\hat Q_p+\hat q_p,\hat m_p x_0 + z \sqrt{\hat
      q_p})
}
\right) \nonumber \\
&&
\left(
\frac{1}{L}\sum_{p=1}^L\int {\cal D}z \int {\rm d}x_0
\left[(1-\rho_0)\delta(x_0) +\rho_0 \phi_0(x_0) \right]
\frac{
1
}
{
1-\rho+\rho g(\hat Q_p+\hat q_p,\hat m_p x_0 + z \sqrt{\hat
      q_p})
}
\right)^{-1}
\\
\ox^{(t+1)} &=&
\frac{1}{\rho}
\left(
\frac{1}{L}\sum_{p=1}^L\int {\cal D}z \int {\rm d}x_0
\left[(1-\rho_0)\delta(x_0) +\rho_0 \phi_0(x_0) \right]
f_a(\hat Q_p+\hat q_p,\hat m_p x_0 + z \sqrt{\hat
      q_p})
\right) \nonumber \\
&&
\left(
\frac{1}{L}\sum_{p=1}^L\int {\cal D}z \int {\rm d}x_0
\left[(1-\rho_0)\delta(x_0) +\rho_0 \phi_0(x_0) \right]
\frac{
g(\hat Q_p+\hat q_p,\hat m_p x_0 + z \sqrt{\hat q_p})
}
{
1-\rho+\rho g(\hat Q_p+\hat q_p,\hat m_p x_0 + z \sqrt{\hat
      q_p})
}
\right)^{-1}
\\
(\sigma^2)^{(t+1)}&=&
\frac{1}{\rho}
\left(
\frac{1}{L}\sum_{p=1}^L\int {\cal D}z \int {\rm d}x_0
\left[(1-\rho_0)\delta(x_0) +\rho_0 \phi_0(x_0) \right]
[f_a(\hat Q_p+\hat q_p,\hat m_p x_0 + z \sqrt{\hat
      q_p})^2+f_c(\hat Q_p+\hat q_p,\hat m_p x_0 + z \sqrt{\hat
      q_p})]\right) \nonumber \\
&&
\left(
\frac{1}{L}\sum_{p=1}^L\int {\cal D}z \int {\rm d}x_0
\left[(1-\rho_0)\delta(x_0) +\rho_0 \phi_0(x_0) \right]
\frac{
g(\hat Q_p+\hat q_p,\hat m_p x_0 + z \sqrt{\hat q_p})
}
{
1-\rho+\rho g(\hat Q_p+\hat q_p,\hat m_p x_0 + z \sqrt{\hat
      q_p})
}
\right)^{-1}-[\overline x^{(t+1)}]^2
\eea
where:
\bea
\hat m_q&=& \frac{1}{L} \sum_p \frac{ J_{pq}\alpha_p}{\Delta +(1/L)\sum_{r=1}^L
   J_{p r}V_r^{(t)}}\ , \label{hatm}\\
\hat q_q&=&\frac{1}{L}\sum_p \frac{ J_{pq}\alpha_p}{[\Delta +(1/L)\sum_{r=1}^L
   J_{p r}V_r^{(t)}]^2}\; \frac{1}{L}\sum_s J_{ps} E_s^{(t)}\ ,\label{hatq}\\
\hat Q_q& =& \frac{1}{L}\sum_p\frac{ \alpha_p
   J_{pq}}{\Delta +(1/L)\sum_{r=1}^L J_{p r}V_r^{(t)}}-\hat q_q = \hat
 m_q - \hat q_q
\, .\label{hatQ}
\eea
The functions $f_a(X,Y)$, $f_c(X,Y)$ were defined in (\ref{f_a}-\ref{f_c}), and the
function $g$ is defined as
\be
   g(X,Y) =  \, \frac{1}{\sqrt{1+X\sigma^2}}\rm{exp}\[[\frac{(Y+\overline x/\sigma^2)^2}{2(X+1/\sigma^2)}-\frac{\overline x^2}{2\sigma^2}\]] .
\ee

\begin{figure}[!ht]
\includegraphics[width=0.6\linewidth]{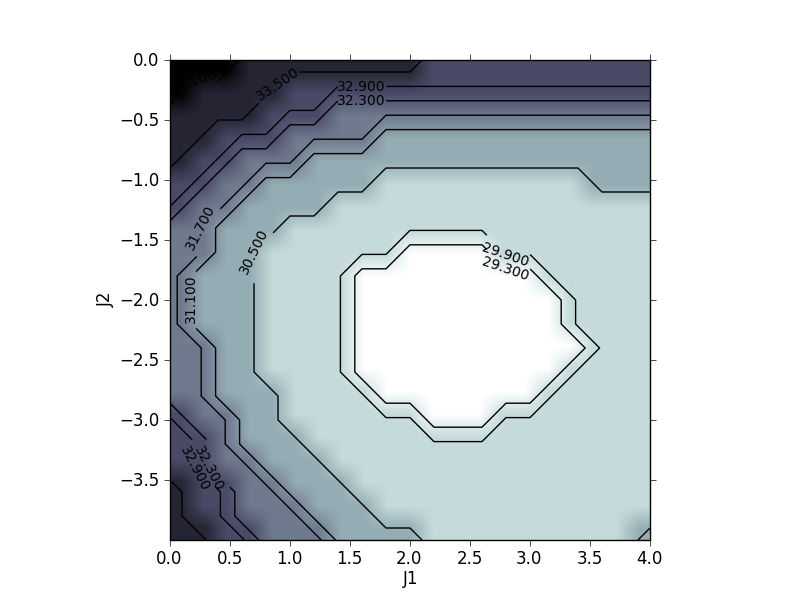}
\caption{\label{J1-J2} (color online): Convergence time of the s-BP
  algorithm with $L=2$
  as a function of $J_1$ and $J_2$ (in $\log$ scale) for Gauss-Bernoulli signal with
  $\rho_0=0.1$.  The color represents the number of iterations such
  that the MSE is smaller than $10^{-8}$. The white color region gives
  the fastest convergence. The
  measurement density is fixed to $\alpha=0.25$.  The parameters in s-BP are
  chosen as $L=2$ and $\alpha_1=0.3$. The axes are in $\log_{10}$ scale.}
\end{figure}

This is the dynamical system that we use in the paper in the noiseless
case ($\Delta=0$) in order to optimize the
values of $\alpha_1$, $J_1$ and $J_2$ . We can estimate the convergence
time of the algorithm as the number of iterations needed in order to
reach the successful fixed point (where all $E_p$ and $V_p$ vanish
within some given accuracy).   Figure~\ref{J1-J2} shows the convergence time
of the algorithm as a function of $J_1$ and $J_2$ for Gauss-Bernoulli signals.

The numerical iteration of this dynamical system is fast. It allows to
obtain the theoretical performance that can be achieved in an
infinite-$N$ system. We have used it in particular to estimate the
values of $L,\alpha_1, J_1, J_2$ that have good performance. For Gauss-Bernoulli
signals, using
optimal choices of $J_1,J_2$, we have found that perfect
reconstruction can be obtained down to the theoretical limit
$\alpha=\rho_0$ by taking $L\to \infty$ (with correction that scale as
$1/L$). Recent rigorous work by Donoho, Javanmard and  Montanari
\cite{DonohoJavanmard11} extends our work and proves our claim that s-BP can reach the optimal threshold asymptotically.

Practical numerical
implementation of s-BP matches  this theoretical performance only when 
the
size of every block is large enough (few hundreds of variables). In
practice, for finite size of the signal, if we want to keep the
block-size reasonable we are hence limited to values of $L$ of several
dozens. Hence in practice we do not quite saturate the threshold
$\alpha=\rho_0$, but exact reconstruction is possible very close to it,
as illustrated in Fig.~2 in the main text, where the values that we
used for the coupling
parameters are listed in Table~\ref{sBEP-param}.

\begin{table}[!ht]
\begin{tabular}{||c|c|c|c|c|c|c||}
  \hline
  $\rho$ & $\alpha$ & $\alpha_1$ & $\alpha'$ & $J_1$ & $J_2$ & $L$ \\
  \hline
  $0.1$ & $0.130$ & $0.3$ & $0.121$ & $1600$ & $1.44$ & $20$ \\
  $0.2$ & $0.227$ & $0.4$ & $0.218$ & $100$ & $0.64$ & $20$ \\
  $0.3$ & $0.328$ & $0.6$ & $0.314$ & $64$ & $0.16$ & $20$ \\
  $0.4$ & $0.426$ & $0.7$ & $0.412$ & $16$ & $0.16$ & $20$ \\
  $0.6$ & $0.624$ & $0.9$ & $0.609$ & $4$ & $0.04$ & $20$ \\
  $0.8$ & $0.816$ & $0.95$ & $0.809$ & $4$ & $0.04$ & $20$ \\
\hline
\end{tabular}
\quad\quad\quad\quad\quad\quad
\begin{tabular}{||c|c|c|c|c|c|c||}
  \hline
  $\rho$ & $\alpha$ & $\alpha_1$ & $\alpha'$ & $J_1$ & $J_2$ & $L$ \\
  \hline
  $0.1$ & $0.150$ & $0.4$ & $0.137$ & $64$ & $0.16$ & $20$ \\
  $0.2$ & $0.250$ & $0.6$ & $0.232$ & $64$ & $0.16$ & $20$ \\
  $0.3$ & $0.349$ & $0.7$ & $0.331$ & $16$ & $0.16$ & $20$ \\
  $0.4$ & $0.441$ & $0.8$ & $0.422$ & $16$ & $0.16$ & $20$ \\
  $0.6$ & $0.630$ & $0.95$ & $0.613$ & $16$ & $0.16$ & $20$ \\
  $0.8$ & $0.820$ & $1$ & $0.811$ & $4$ & $0.04$ & $20$ \\
 \hline
\end{tabular}
\caption{Parameters used for the s-BP reconstruction of the Gaussian signal (left) and of the binary signal (right). \label{sBEP-param}}
\end{table}

In Fig.~3, we also presented the result of the s-BP reconstruction of
the Gaussian signal of density $\rho_0=0.4$ for different values of
$L$. We have observed empirically that the result is rather robust to
choices of $J_1, J_2$ and $\alpha_1$. In this case, in order to
demonstrate that very different choices give seeding matrices which
are efficient, we used $\alpha_1=0.7$, and then $J_1=1043$ and
$J_2=10^{-4}$ for $L=2$,$J_1=631$ and $J_2=0.1$ for $L=5$, $J_1=158$
and $J_2=4$ for $L=10$ and $J_1=1000$ and $J_2=1$ for $L=20$. One sees
that a common aspect to all these choices is a large ratio $J_1/J_2$.
Empirically, this seems to be important in order to ensure a short
convergence time. A more detailed study of convergence time of the
dynamical system will be necessary in order to give some more
systematic rules for choosing the couplings. This is left for future
work.

Even though our theoretical study of the seeded BP was performed on an
example of a specific signal distribution, the examples presented in
Figs.~1 and 2 show that the performance of the algorithm is robust and
also applies to images which are not drawn from that signal
distribution.

\section{Phase diagram in the variables used by \cite{Donoho:2005wq}}
\begin{figure}[h]
  \begin{center}
    \includegraphics[scale=0.5]{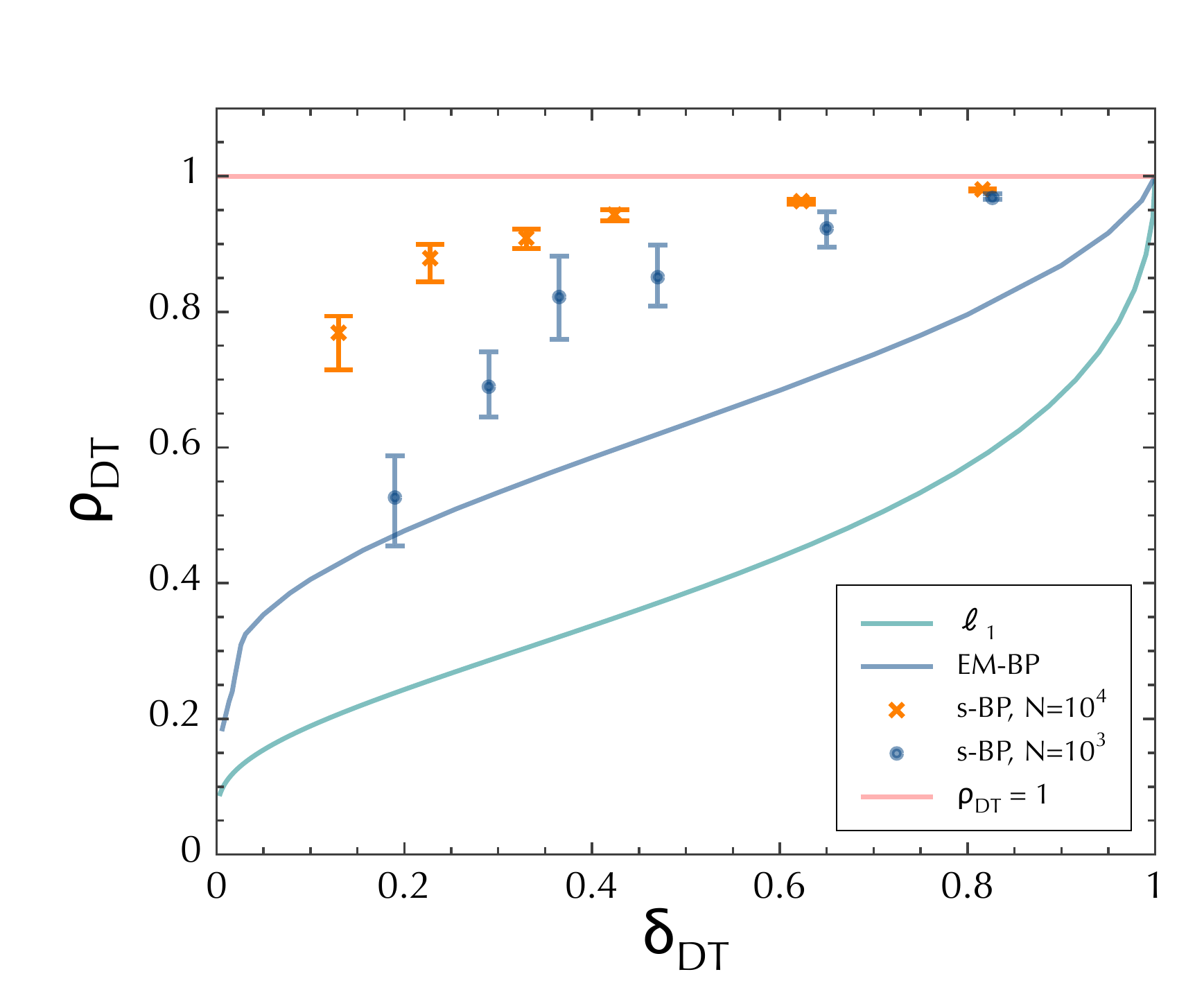}
    \includegraphics[scale=0.5]{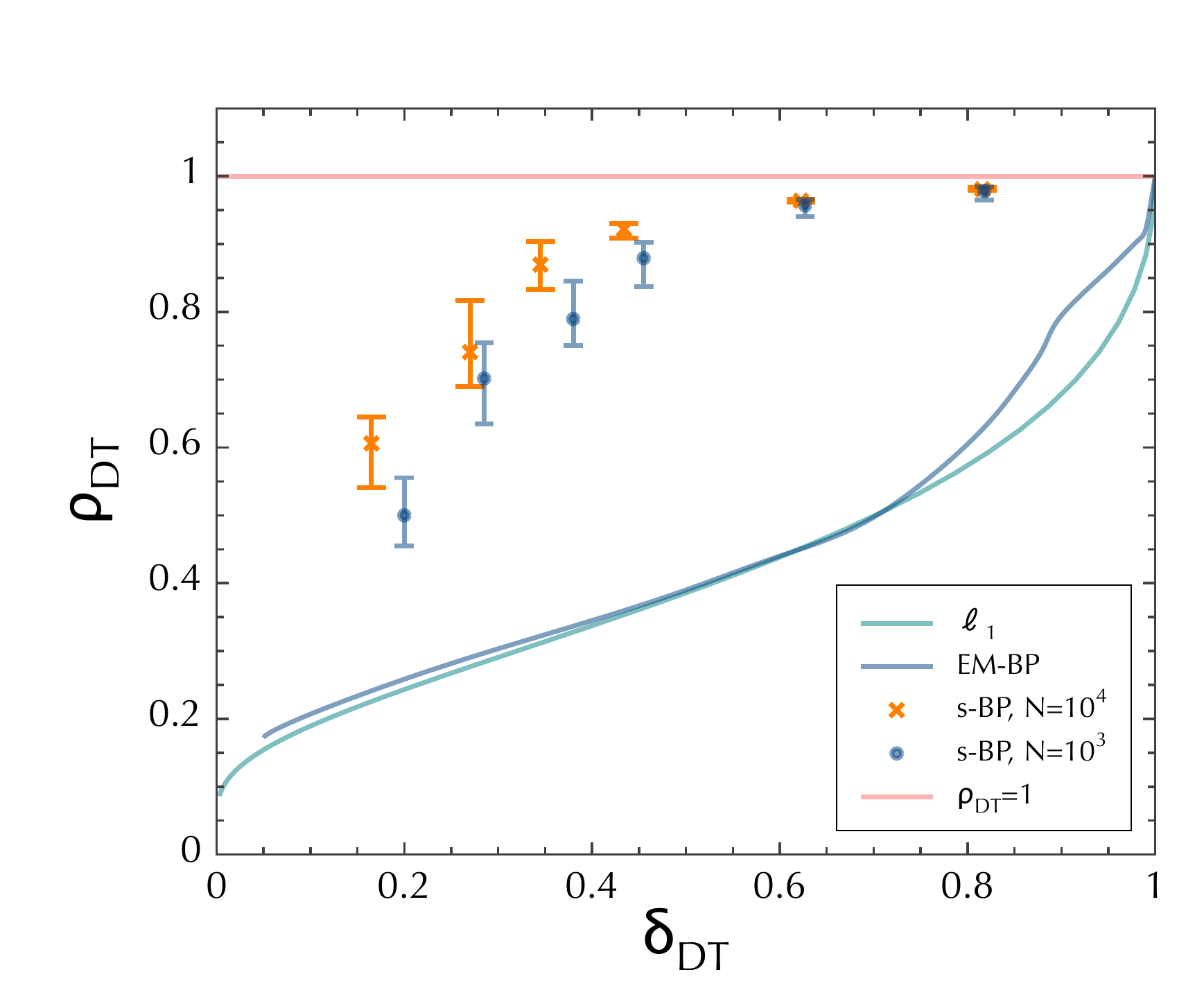}
      \caption{(color online): Same data as in Fig.~2 in the main paper, but using the
         convention of  \cite{Donoho:2005wq}. The phase diagrams is
         now plot as a function of the under-sampling ratio
         $\rho_{\emph DT}=K/M=\rho/\alpha$ and of the over-sampling ratio
         $\delta_{\emph DT}=M/N=\alpha$.}
  \label{fig:phase_diag_DT}
\end{center}
\end{figure}

We show in  Fig.~\ref{fig:phase_diag_DT} the phase diagram
in the convention used by \cite{Donoho:2005wq}, which might be more
convenient for some readers.

\section{Details on the phantom and Lena examples}
\label{details}
In this section, we give a detailed description of the way we have
produced the two examples of reconstruction in Fig.~1.
It is important to stress that this figure is intended to be
an illustration of the s-BP reconstruction algorithm. As such, we
have used elementary protocols to produce true $K$-sparse signals and
have not tried to optimize the sparsity nor to use the best possible
compression algorithm; instead, we have limited ourselves to the
simplest Haar wavelet transform, to make the exact reconstruction and the
comparison between the different approaches more transparent.

The Shepp-Logan example is a $128^2$ picture that has been generated
using the Matlab implementation. The Lena picture is a $128^2$ crop of
the $512^2$ gray version of the standard test image.
 In the first
case, we have worked with the sparse one-step Haar transform of the picture,
while in the second one, we have worked with a modified picture where
we have kept the $24$ percent of largest (in absolute value) coefficients
of the two-step Haar transform, while putting all others to zero.
The datasets of the two images are available online \cite{noteASPICS}.
Compressed sensing here is done as follows: The original image is a
vector $\bo$ of $N=L^2$ pixels. The unknown vector $\bx=\bW \bo$ are
the projections of the original image on a basis of one- or two-steps
Haar wavelets. It is sparse by construction. We generate a matrix
$\bF$ as described above, and construct $\bG =\bF \bW$. The
measurements are obtained by $\by= \bG \bo$, and the linear system for
which one does reconstruction is $\by=\bF \bx$. Once $\bx$ has been
found, the original image is obtained from $\bo =\bW^{-1} \bx$. We
used EM-BP and s-BP with a Gauss-Bernoulli $P(\bx)$.

On the algorithmic side, the EM-BP and $\ell_1$ experiments were run with
the same full gaussian random measurement matrices. The minimization of the $\ell_1$ norm was done
using the $\ell_1$-magic tool for Matlab, which is a free
implementation that can be download at {\bf
  http://users.ece.gatech.edu/\textasciitilde justin/l1magic/}, using the lowest
possible tolerance such that the algorithm outputs a solution. The
coupling parameters of s-BP  are given in
Table~\ref{phantom-param}. A small damping was used in each
iterations, we mixed the old and new messages, keeping $20\%$ of the
messages from the previous iteration. Moreover, since for both Lena
and the phantom, the components of the signal are correlated, we have
permuted randomly the 
columns
of the sensing matrix $F$. This allows to
avoid the (dangerous) situation where a given block contains only zero
signal components.

Note that the number of iterations needed by the s-BP procedure to
find the solution is moderate. The s-BP algorithm coded in Matlab
finds the image in few seconds. For instance, the Lena picture at
$\alpha=0.5$ requires about $500$ iterations and about $30$ seconds on
a standard laptop. Even for the most difficult case of Lena at
$\alpha=0.3$, we need around $2000$ iterations, which took only about
$2$ minutes. In fact, s-BP (coded in Matlab) is
much faster than the Matlab implementation of $\ell_1$-magic on the
same machine. We report the MSE for all these reconstruction protocols
in Table \ref{phantom-MSE}.

\begin{table}[!ht]
\begin{tabular}{||c|c|c|c|c|c||}
  \hline
  $\alpha$ & $\alpha_1$ & $\alpha'$ & $J_1$ & $J_2$ & $L$ \\
  \hline
  $0.5$ & $0.6$ & $0.495$ & $20$ & $0.2$ & $45$ \\
  $0.4$ & $0.6$ & $0.395$ & $20$ & $0.2$ & $45$ \\
  $0.3$ & $0.6$ & $0.295$ & $20$ & $0.2$ & $45$ \\
  $0.2$ & $0.6$ & $0.195$ & $20$ & $0.2$ & $45$ \\
  $0.1$ & $0.3$ & $0.095$ & $1$ & $1$ & $30$ \\
\hline
\end{tabular}
\quad\quad\quad\quad\quad\quad
\begin{tabular}{||c|c|c|c|c|c||}
  \hline
  $\alpha$ & $\alpha_1$ & $\alpha'$ & $J_1$ & $J_2$ & $L$ \\
  \hline
  $0.6$ & $0.8$ & $0.585$ & $20$ & $0.1$ & $30$ \\
  $0.5$ & $0.8$ & $0.485$ & $20$ & $0.1$ & $30$ \\
  $0.4$ & $0.8$ & $0.385$ & $20$ & $0.1$ & $30$ \\
  $0.3$ & $0.8$ & $0.285$ & $20$ & $0.1$ & $30$ \\
  $0.2$ & $0.5$ & $0.195$ & $1$ & $1$ & $30$ \\
 \hline
\end{tabular}
\caption{Parameters used for the s-BP reconstruction of the Shepp-Logan
  phantom image (left) and of the Lena image (right).
 \label{phantom-param}}
\end{table}

\begin{table}[!ht]
\begin{tabular}{||r|c|c|c|c|c||}
  \hline
   & $\alpha=0.5$ & $\alpha=0.4$ & $\alpha=0.3$ & $\alpha=0.2$ & $\alpha=0.1$ \\
  \hline
$\ell_1$ & $0$ & $0.0055$ & $0.0189$ & $0.0315$ & $0.0537$ \\
BP& $0$ & $0$ & $0.0213$ & $0.025$ & $0.0489$ \\
s-BP& $0$ & $0$ & $0$ & $0$ & $0.0412$ \\
\hline
\end{tabular}
\quad\quad
\begin{tabular}{||c|c|c|c|c|c||}
  \hline
 & $\alpha=0.6$ & $\alpha=0.5$ & $\alpha=0.4$ & $\alpha=0.3$ & $\alpha=0.2$ \\
  \hline
$\ell_1$ &0 & $4.48\, . 10^{-4}$ & $0.0015$ & $0.0059$ & $0.0928$ \\
BP&0 & $4.94\, . 10^{-4}$ & $0.0014$ & $0.0024$ & $0.0038$ \\
s-BP&  $0$  & $0$ & $0$ & $0$ & $0.0038$  \\
 \hline
\end{tabular}
\caption{Mean-squared error obtained after reconstruction for the Shepp-Logan phantom (left)
  and for Lena (right) with $\ell_1$ minimization, BP and s-BP.\label{phantom-MSE}}
\end{table}

\begin{figure}[!ht]

\includegraphics[width=0.47\linewidth]{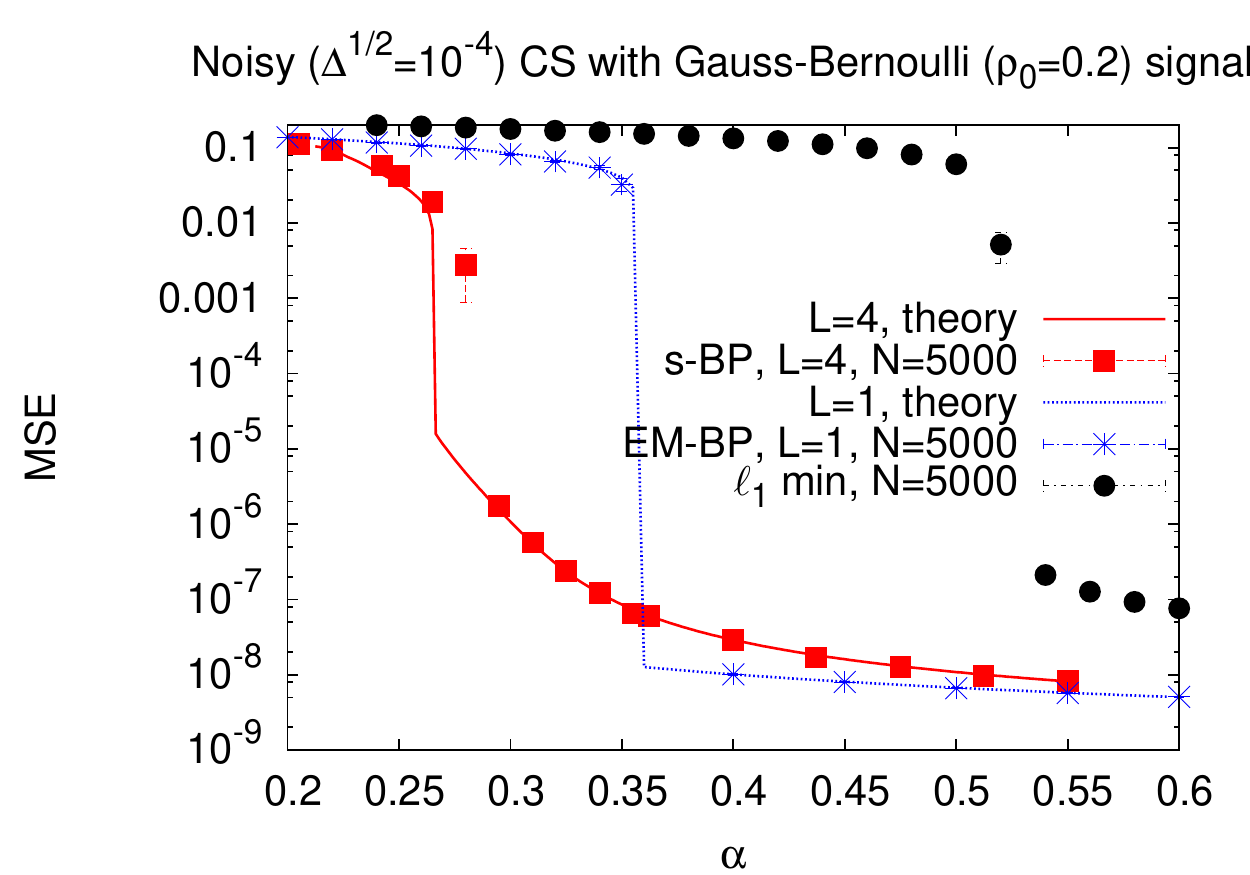}
\includegraphics[width=0.47\linewidth]{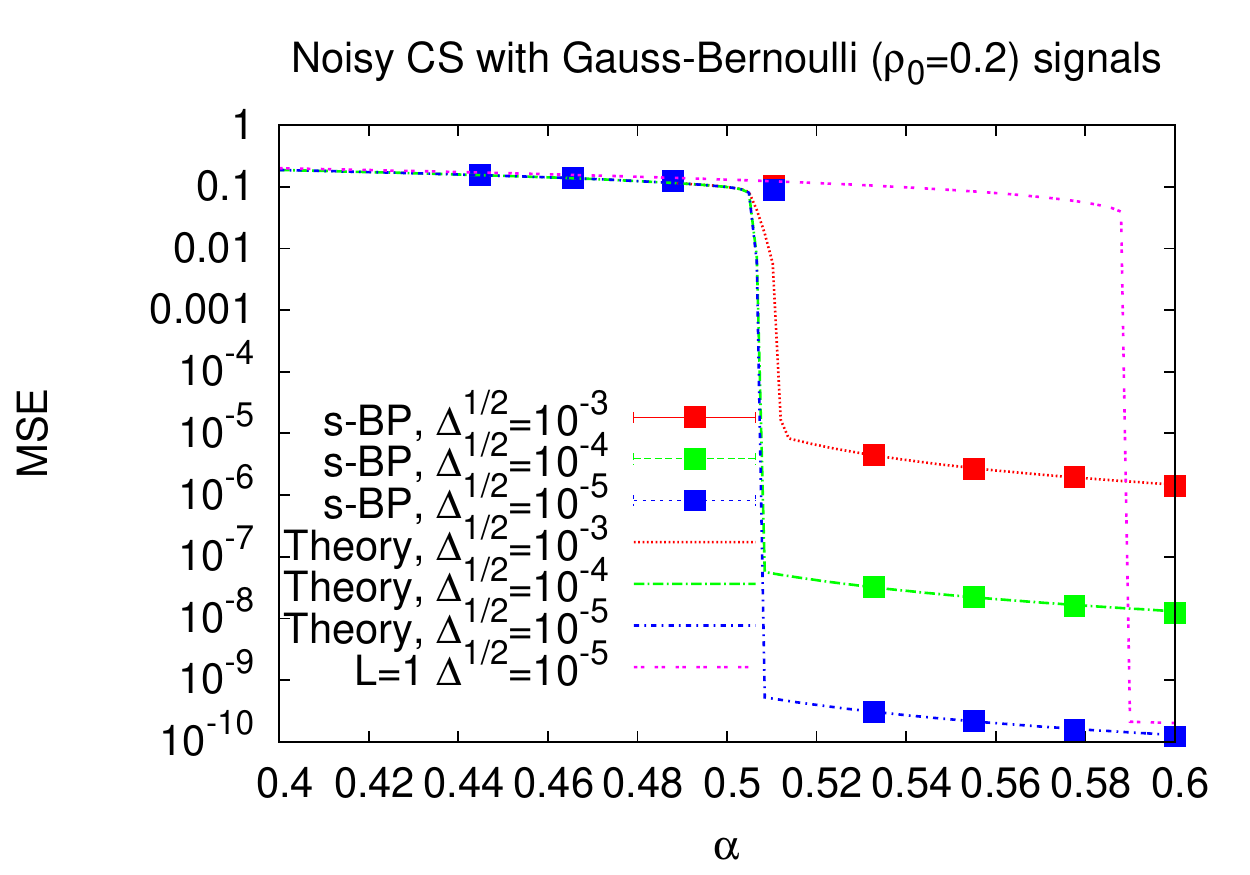}
\caption{\label{error_noise} (color online): Mean-squared error as a
  function of $\alpha$ for different values of the measurement
  noise. {\bf Left:}  Gauss-Bernoulli
  signal with $\rho_0=0.2$, and measurement noise with standard
  deviation $\sqrt{\Delta}=10^{-4}$.
The EM-BP ($L=1$) and s-BP strategy ($L=4,J_1=20,J=0.1,\alpha_1=0.4$) are able to
  perform a very good reconstruction up to much lower value of
  $\alpha$ than the $\ell_1$ procedure. Below a critical value of
  $\alpha$, these algorithms show a first order phase transition to a
  regime with  much larger MSE. {\bf Right:}  Gauss-Bernoulli
  signal with $\rho_0=0.4$, with a noise with standard
  deviation $\sqrt{\Delta}=10^{-3},10^{-4},10^{-5}$. s-BP, with $L=9,
  J_1=30,J_2=8,\alpha_1=0.8$ decodes very well for all these values of
  noises. In this case, $\ell_1$ is unable to reconstruct for all $\alpha<0.75$, well
  outside the range of this plot. }
\end{figure}
\section{Performance of the algorithm in the presence of measurement  noise}
A systematic study of our algorithm for the case of noisy measurements
can be performed using the replica analysis, but goes beyond the scope
of the present work. In this section we however want to point out two
important facts: 1) the modification of our algorithm to take into
account the noise is straightforward, 2) the results that we have
obtained are robust to the presence of a small amount of noise.
As shown in section~\ref{BP_der}, the probability of $\bx$ after the
measurement of $\by $ is given by \be \hat P(\bx) = \frac{1}{Z}
\prod_{i=1}^N \left[ (1-\rho) \delta(x_i) + \rho \phi(x_i) \right]
\prod_{\mu=1}^M \frac{1}{\sqrt{2\pi \Delta_\mu}}
e^{-\frac{1}{2\Delta_\mu}(y_\mu - \sum_{i=1}^N F_{\mu i}x_i)^2} \ee
$\Delta_{\mu}$ is the variance of the Gaussian noise in measurement
$\mu$. For simplicity, we consider that the noise is homogeneous,
i.e., $\Delta_{\mu}=\Delta$, for all $\mu$, and discuss the result in
unit of standard deviation $\sqrt{\Delta}$. The AMP-form of the
message passing including noise have already been given in
section~\ref{sec:TAP}. The variance of the noise, $\Delta$, can also
be learned via expectation maximization approach, which reads: \be
\Delta =
\frac{\sum_{\mu}\frac{(y_{\mu}-\omega_{\mu})^2}{(1+\frac{1}{\Delta}
    \gamma_{\mu})^2}}{\sum_{\mu}\frac{1}{1+\frac{1}{\Delta}\gamma_{\mu}}}
\, .
\label{delta_learn}
 \ee
 The dynamical system describing the density evolution has been
 written in Sect.~\ref{replica_1D}. It just needs to be complemented by
 the iteration on $\Delta$ according to Eq.~(\ref{delta_learn}).
 We have used it to study the evolution of MSE in the case
 where both $P(x)$ and signal distribution are Gauss-Bernoulli with
 density $\rho$, zero mean and unit variance for the full measurement
 matrix and for the seeding one. Figure~\ref{error_noise} shows the MSE as
 a function of $\alpha$ for a given $\Delta$, using our algorithm, and
 the $\ell_1$ minimization for comparison. The analytical results
 obtained by the study of the dynamical system are compared to
 numerical simulations, and agree very well.

\bibliographystyle{nature}
\bibliography{refs}

\end{document}